\documentclass[acmlarge,screen,nonacm]{acmart}
\AtBeginDocument{%
 }

\usepackage{adjustbox}
\usepackage{multirow} 
\usepackage{multirow}
\usepackage[table,xcdraw]{xcolor}
\usepackage[normalem]{ulem}
\usepackage{arydshln}
\usepackage{booktabs}
\usepackage{multirow}

\newif\ifrevmark
\revmarktrue
\DeclareRobustCommand{\rev}[1]{\ifrevmark\textcolor{black}{#1}\else#1\fi}

\usepackage{amssymb}
\newcommand{\faCheck}{\ensuremath{\checkmark}}
\newcommand{\faTimes}{\ensuremath{\times}}

\renewcommand\footnotetextcopyrightpermission[1]{} 
\setcopyright{none}
\begin{document}

\title{Person Parametric Physics-informed Representation for mmWave-based Human Pose Estimation}

\author{Shuntian Zheng}
\email{shuntian.zheng@warwick.ac.uk}
\affiliation{%
  \institution{Department of Computer Science, University of Warwick}
  \city{Coventry}
  \state{West Midlands}
  \country{United Kingdom}
}

\author{Jiaqi Li}
\email{jiaqi.li.16@warwick.ac.uk}
\affiliation{%
  \institution{University of Warwick}
  \city{Coventry}
  \state{West Midlands}
  \country{United Kingdom}
}

\author{Guangming Wang}
\email{gw462@cam.ac.uk}
\affiliation{%
  \institution{University of Cambridge}
  \city{Cambridge}
  \country{United Kingdom}
}

\author{Minzhe Ni}
\email{Minzhe.Ni@warwick.ac.uk}
\affiliation{%
  \institution{Department of Computer Science, University of Warwick}
  \city{Coventry}
  \state{West Midlands}
  \country{United Kingdom}
}

\author{Arnad Palit}
\email{a.palit.1@warwick.ac.uk}
\affiliation{%
  \institution{Warwick Manufacturing Group, University of Warwick}
  \city{Coventry}
  \state{West Midlands}
  \country{United Kingdom}
}

\author{Giovanni Montana}
\email{g.montana@warwick.ac.uk}
\affiliation{%
  \institution{Department of Statistics, University of Warwick}
  \city{Coventry}
  \state{West Midlands}
  \country{United Kingdom}
}

\author{Yu Guan}
\email{Yu.Guan@warwick.ac.uk}
\affiliation{%
  \institution{Department of Computer Science, University of Warwick}
  \city{Coventry}
  \state{West Midlands}
  \country{United Kingdom}
}

\renewcommand{\shortauthors}{Zheng et al.}


\begin{abstract}

Millimeter-wave (mmWave) radar enables privacy-preserving, illumination-invariant Human Pose Estimation (HPE). However, current mmWave-based HPE systems face a signal-noise dilemma: Heatmaps retain human reflections but embed environmental clutter, while Point Clouds (PC) suppress noise through aggressive thresholding but discard informative human reflections, limiting robustness across environments and radar configurations.
\rev{To address this intrinsic bottleneck, we introduce Person Parametric Physics-informed Representation (PPPR), a physics-informed parametric intermediate representation that replaces purely signal-level encodings with human-centric parameterization}. 
PPPR models each human joint as a Gaussian primitive encoding both kinematic properties, which include position, velocity, orientation, and electromagnetic properties, which include scattering intensity and Doppler signature. These parameters enable optimization through a dual-constraint process: kinematic objectives enforce biomechanical consistency to suppress spatial artifacts, while electromagnetic objectives ensure adherence to mmWave propagation physics, decoupling input representations from non-human noise. Experiments across three mmWave-based HPE datasets with four HPE models demonstrate that replacing conventional inputs with PPPR consistently yields substantial accuracy gains. Furthermore, cross-scenes and cross-datasets experiments confirm PPPR’s noise decoupling capability: models trained with PPPR maintain stable performance across diverse furniture arrangements and different radar chipsets, demonstrating its promising generalization capability in the challenging cross-dataset settings. Code will be released upon publication.
\end{abstract}




\maketitle

\section{Introduction}
\label{sec:Introduction}

Human Pose Estimation (HPE) underpins a wide range of ubiquitous sensing applications such as elderly fall detection, gesture-based smart home control, and in-home rehabilitation guidance~\cite{zhao2018through,zhang2024single,liu2024pmtrack,engel2025advanced}. 
Traditional camera-based systems capture rich visual signals but are vulnerable to illumination variations, occlusions, and privacy constraints~\cite{zhao2018rf,guhr2020privacy,alshehri2022exploring}. 
Wearable devices, while unaffected by the environment, require user compliance for charging, placement, and comfort, which limits sustained adoption~\cite{liu2025umotion,salehzadeh2024wearable,toh2023usability}.  
Millimeter-wave (mmWave) radar has recently emerged as a promising alternative that preserves privacy and operates robustly across lighting and non-contact conditions~\cite{chang2020spatial,sengupta2022mmpose, mei2024mmspyvr, palipana2021pantomime}.  

Although mmWave-based HPE systems provide these advantages, their effectiveness remains limited by an intrinsic signal–noise dilemma in the input data.  
To understand this challenge, it is essential to first examine the unique noise characteristics of mmWave signals.  
Radar reflections inherently combine weak human-related signals with overwhelming non-human components originating from (1) environmental noise, such as multipath interference from furniture and walls \cite{zhang2025breaking}, and (2) radar hardware noise, such as ADC quantization and internal signal distortion \cite{ren2024noncontact}.  
These two sources are inseparably coupled with human reflections in mmWave-based HPE models' inputs \cite{han20234d}, making it difficult to simultaneously preserve complete human information and suppress irrelevant noise, severely constraining mmWave-based HPE systems' performance and stability.

Two primary input formats dominate current mmWave-based HPE systems: Heatmap and Point Cloud (PC), yet both are affected by environmental and radar hardware noises.
The Heatmap, obtained through Fast Fourier Transform (FFT)-based frequency-domain extraction, preserves nearly all spatial perturbations in a scene, retaining abundant motion cues but embedding massive noise.  
Consequently, true human-related reflections often occupy only a small fraction of the total captured energy and are easily obscured by clutter.  
On the other hand, PC, generated via Constant False Alarm Rate (CFAR) thresholding~\cite{rohling1983radar,cao2022joint}, improves the signal-to-noise ratio by discarding low-intensity reflections.  
However, this aggressive thresholding causes irreversible sparsification: informative reflections related to subtle joint movements are frequently removed, while occasional environmental or radar-induced high-intensity points may be mistakenly retained \cite{wang2023human, cai2023millipcd,liu2024view,qian20203d}.  
As a result, Heatmaps suffer from high information retention but excessive noise, while PC achieves low noise but at the cost of losing critical information.  
Neither can simultaneously achieve maximum human information preservation and minimum noise interference—a dilemma that originates from the input design itself rather than mmWave-based HPE modeling.

\rev{To resolve this input-level challenge, we propose Person Parametric Physics-informed Representation (PPPR), a physics-informed \emph{parametric intermediate representation} designed to emphasize human-related components while suppressing non-human clutter.}  
Inspired by Gaussian Splatting~\cite{Wu_2024_CVPR}, PPPR models each human joint as a Gaussian primitive that encodes both kinematic and electromagnetic characteristics.  
The kinematic parameters describe position, velocity, and orientation, while the electromagnetic parameters capture radar scattering intensity and Doppler behavior~\cite{iovescu2020fundamentals}.  
Unlike conventional data-driven noise suppression approaches, PPPR introduces two complementary physics-informed optimization objectives:  
(1) Kinematic objectives enforce universal biomechanical consistency—skeletal topology, joint angle limits, and limb length ratios—ensuring that retained features align with human motion constraints;  
(2) Electromagnetic objectives ensure physical consistency with mmWave signal propagation, including Doppler relationships and antenna-array responses, thereby suppressing artifacts linked to radar hardware or signal-processing irregularities.  
Together, these two constraints enable PPPR to achieve the ideal trade-off: maximizing the preservation of human-related information while minimizing both environmental and radar noise.


To construct and optimize PPPR efficiently, we introduce MmWave Human Parameterization (MHP), a differentiable pipeline that initializes and refines PPPR parameters.  
MHP begins with initialization, extracting coarse joint regions and signal flow directions from the Heatmap to seed PPPR parameters.  
Then, in the optimization stage, PPPR parameters are iteratively refined according to kinematic and electromagnetic objectives.  
To prevent overfitting toward either constraint, MHP includes a radar simulation module that reconstructs a synthetic Heatmap from PPPR via a differentiable radar forward pipeline \cite{iovescu2020fundamentals}.  
By comparing the simulated and original Heatmaps, MHP identifies and excludes false human signals.  

\begin{figure*}[h]
  \centering
  \includegraphics[width=0.75\linewidth]{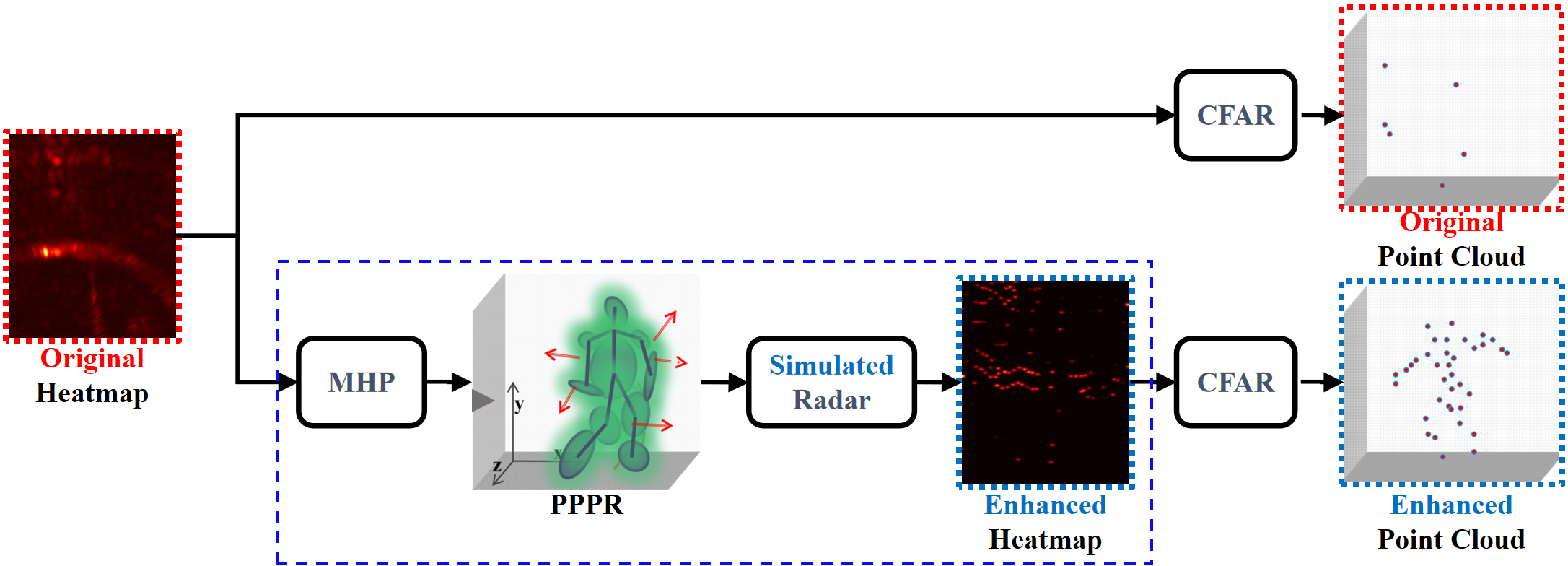}
  \caption{Person Parametric Physics-informed Representation (PPPR) processing 
flow. Input: Heatmap; Output: PPPR parameters; Additional Output: PPPR-enhanced Heatmap and PPPR-enhanced PC, for adaptability on existing Heatmap-based or PC-based HPE models.}
  \label{Heatmap-PC-PPPR} 
\end{figure*}

The optimized PPPR can serve as a direct input for mmWave-based HPE models or be converted back into synthetic Heatmaps or PC, both retaining PPPR’s denoised and feature-complete characteristics.  
These regenerated inputs, as highlighted in blue in Fig.~\ref{Heatmap-PC-PPPR}, preserve more human skeletal details with substantially less non-human noise than the original inputs indicated in red. 
Subsequent experiments systematically evaluate three input configurations: (1) original Heatmap \cite{yataka2024retr,lee2023hupr}/PC \cite{fan2024diffusion}, (2) PPPR, and (3) PPPR-reconstructed Heatmap/PC, demonstrating that even the regenerated representations retain PPPR’s advantages in signal fidelity and noise suppression.

This work makes the following key contributions:
\begin{itemize}

    \item \textbf{\rev{A physics-informed parametric intermediate representation for mmWave-based HPE.}}  
    \rev{We identify the signal–noise imbalance that constrains existing input formats and propose PPPR, a parametric intermediate representation that preserves human motion cues while suppressing non-human clutter through physics-grounded constraints.}   
    \item \textbf{Physics-informed dual-objective optimization.}  
    We introduce MmWave Human Parameterization (MHP), a complete pipeline that optimizes PPPR using kinematic and electromagnetic objectives grounded in physical principles, ensuring accurate, physically consistent, and noise-resistant representation.
    \item \textbf{Comprehensive experimental validation.}  
    Based on four representative HPE models and three public datasets (MMVR \cite{rahman2024mmvr}, HuPR \cite{lee2023hupr}, and XRF55 \cite{wang2024xrf55}) collected under diverse environments and radar configurations, we evaluate the performance of PPPR under various testing modes.  
    The results show that when using PPPR as input, all four models achieve higher accuracy and greater stability compared to their counterparts using traditional Heatmap or PC.  
    These consistent improvements across different models and datasets indicate that PPPR effectively mitigates both environmental and radar-induced noise while preserving complete human motion information.  
    Moreover, in challenging multi-subject scenarios, PPPR maintains stable accuracy and clear person-wise separation, demonstrating its potential in complex radar sensing conditions.
\end{itemize}

\section{Preliminaries}

\subsection{Millimeter Wave Radar Processing Pipeline}

\begin{figure*}[h]
  \centering
  \includegraphics[width=0.85\linewidth]{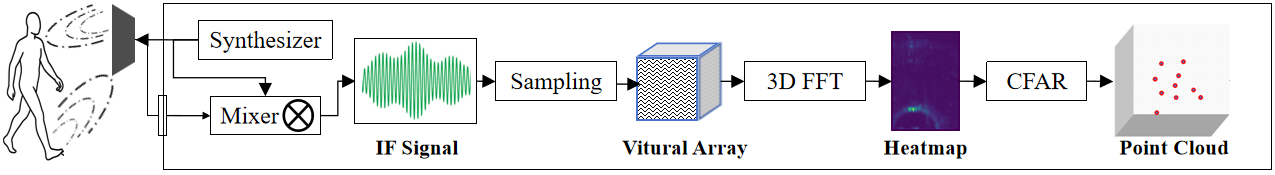}
  \caption{Millimeter wave radar signal processing pipeline: FMCW chirp transmission, three-stage FFT processing (Range, Doppler, Angle), and CFAR-based Point Cloud extraction.}
  \label{radar_pipeline} 
\end{figure*}

Millimeter wave (mmWave) radar operates on the principle of electromagnetic wave transmission and reflection analysis, utilizing frequency-modulated continuous wave (FMCW) technology to transform raw electromagnetic signals into structured data representations suitable for HPE systems \cite{iovescu2020fundamentals}. The radar transmits chirp signals characterized by linearly increasing frequency over time, defined by start frequency $f_s$, bandwidth $B$, and chirp duration $T_c$. When transmitted signals encounter objects in the environment, reflections are captured by receive antennas, creating time-delayed versions of the original chirp with delay $\tau = 2r/c$, where $r$ represents object range and $c$ denotes the speed of light.

The signal processing pipeline consists of three cascaded FFT stages \cite{iovescu2020fundamentals} that extract range, velocity, and angular information sequentially. \textbf{(1) Range FFT:} Transmitted and received signals are mixed to generate intermediate frequency (IF) signals. For objects at range $r$, this produces beat frequencies $f_{\text{beat}} = S \cdot 2r/c$, where $S = B/T_c$ represents the chirp slope. The first FFT transforms IF signals into the frequency domain, where peaks correspond to target ranges. \textbf{(2) Doppler FFT:} The second FFT operates across multiple chirps within a frame, extracting velocity through Doppler frequency shifts \cite{chen2006micro}: $\Delta f_d = 2v_r/\lambda$, where $v_r$ is radial velocity and $\lambda$ is wavelength. \textbf{(3) Angle FFT:} The third FFT processes phase differences across multiple receive antennas to obtain angular localization. 

In typical mmWave radar designs, the transmit–receive (Tx–Rx) array adopts an L-shaped layout on the printed circuit board \cite{iovescu2020fundamentals,rahman2024mmvr}: horizontal antenna elements with spacing $l_{\mathrm{az}}$ measure azimuth phase differences $\Delta\Phi_{\mathrm{az}}$, while vertical elements with spacing $l_{\mathrm{el}}$ measure elevation phase differences $\Delta\Phi_{\mathrm{el}}$. The angles are computed as $\theta_{\mathrm{az}}=\sin^{-1}(\lambda \Delta\Phi_{\mathrm{az}}/2\pi l_{\mathrm{az}})$ and $\theta_{\mathrm{el}}=\sin^{-1}(\lambda \Delta\Phi_{\mathrm{el}}/2\pi l_{\mathrm{el}})$ respectively. This three-stage processing produces a range–angle–Doppler data cube, which is projected into a Heatmap representing spatial intensity distributions. Additionally, unlike optical signals, radar signals undergo round-trip propagation \cite{piersanti2012millimeter}, resulting in $r^{-4}$ power attenuation.

Point Cloud (PC) is derived from the Heatmap through the Constant False Alarm Rate (CFAR) algorithm \cite{rohling1983radar}. CFAR employs adaptive thresholding to suppress clutter: for each cell, a threshold is computed from surrounding reference cells' noise statistics, and peaks exceeding this threshold are retained as detections. Each detected peak is converted to a 3D coordinate $(x, y, z)$ through geometric transformation from range–angle coordinates. This discretization process reduces data dimensionality but introduces sparsity, as continuous signal distributions are reduced to discrete point representations \cite{yang2025spatial}.

Heatmap and PC exhibit substantial differences in physical semantics. The Heatmap preserves comprehensive signal information but inherently contains environmental noise from multipath reflections and radar processing artifacts. PC achieves an improved signal-to-noise ratio through CFAR filtering but exhibits information sparsity due to aggressive thresholding that discards weak but potentially informative human reflections.

\subsection{Gaussian Splatting: Preventing Prior Overfitting via Differentiable Re-rendering}

Gaussian Splatting (GS) \cite{Wu_2024_CVPR}  has demonstrated the efficacy of parametric primitive representations in computer vision, modeling 3D scenes through collections of differentiable Gaussian functions. Each Gaussian primitive is characterized by position $\mathbf{p} \in \mathbb{R}^3$, covariance $\mathbf{\Sigma} \in \mathbb{R}^{3 \times 3}$, color $c$, and opacity $\alpha$, enabling explicit spatial editing while maintaining rendering efficiency.

The critical innovation preventing parameter overfitting lies in GS's differentiable rendering pipeline.
Rather than optimizing parameters solely under geometric priors, GS reconstructs observations from optimized parameters through alpha-blended compositing.
This re-rendering process is formalized as:
\begin{equation}
\small
C(\mathbf{u}) = \sum_{i=1}^{N} c_i \alpha_i \prod_{j=1}^{i-1}(1-\alpha_j) \cdot \exp\left(-\frac{1}{2}(\mathbf{u} - \mathbf{u}_i)^T \mathbf{\Sigma}_i^{2D^{-1}} (\mathbf{u} - \mathbf{u}_i) \right),
\label{eq:gs_rendering}
\end{equation}
where $C(\mathbf{u})$ denotes the rendered color at pixel $\mathbf{u}$, $\mathbf{u}_i$ is the projected centroid of the $i$-th Gaussian, and $\mathbf{\Sigma}_i^{2D}$ represents the covariance projected onto the image plane.
Here, $N$ indicates the total number of Gaussian primitives contributing to the rendered scene, each acting as a differentiable volumetric element in the reconstruction.
By comparing synthetic images $C(\mathbf{u})$ against observed photometry, optimization is grounded in photometric consistency rather than relying solely on geometric priors that may generate physically implausible configurations.
These constraints ensure parameters remain faithful to physical observations while satisfying structural coherence.

\textbf{Extension toward PPPR-based modeling.} 
Inspired by the differentiable re-rendering mechanism in Gaussian Splatting, we extend this concept to radar-based human pose reconstruction in MHP. 
Unlike visual rendering, mmWave radar systems require jointly modeling electromagnetic propagation and human kinematics.

\section{Related work}

\subsection{Denoising in mmWave-based HPE}

Millimeter-wave (mmWave) signals contain a mixture of human reflections and diverse noise sources, including multipath interference from furniture, as well as radar-induced distortions~\cite{rahman2024mmvr,yataka2024retr,lee2023hupr,fan2024diffusion}.  
Therefore, existing Human Pose Estimation (HPE) depends critically on how these noisy signals are represented before the learning model.  
Current mmWave-based HPE datasets typically provide the Heatmap or PC as input (Table 1). As a result, existing approaches adopt two dominant denoising strategies (Table 2): implicit denoising in Heatmap-based pipelines and explicit denoising in Point Cloud (PC)-based pipelines.

Heatmap-based models~\cite{cui2021real,wu2024mmhpe,zhang2021comprehensive,rahman2024mmvr} retain dense spatial-temporal intensity maps that encode both human motion and background reflections.  
These HPE models rely on neural architectures to implicitly distinguish human signals from environmental noise through learned filters.  
Common frameworks include convolutional networks~\cite{cui2021real}, multi-task feature decoupling~\cite{wu2024mmhpe,zhang2021comprehensive}, and attention mechanisms~\cite{rahman2024mmvr}. 
However, because the raw input preserves nearly all reflections, deep networks inevitably learn correlations tied to scene geometry, furniture layout, or radar hardware characteristics~\cite{yataka2024retr,zhu2024probradarm3f}.  
While these correlations may aid training performance, they lack physical grounding and thus do not generalize to new environments or radar hardware conditions.  
The key limitation is conceptual: learning-based denoising operates as a statistical filter rather than a physically interpretable suppression mechanism, making it difficult to isolate universal human motion cues from noise distributions that change with context.

PC-based models~\cite{sengupta2020mm,kini2024transhupr,fan2024diffusion, feng20243d} perform denoising at the preprocessing stage by applying Constant False Alarm Rate (CFAR) thresholding~\cite{rohling1983radar} to the raw Heatmap.  
CFAR adaptively filters low-intensity reflections, retaining only strong returns as 3D point coordinates with improved signal-to-noise ratio.  
This explicit approach effectively removes large portions of clutter but also introduces irreversible information loss.  
Weak yet informative reflections—particularly from small or fast-moving joints—are often discarded alongside environmental noise.  
Moreover, the CFAR threshold depends on local statistics influenced by the scene’s reflective surfaces and radar-specific detection characteristics, producing input sparsity patterns that vary with both environment and radar configuration.  
Subsequent models attempt to recover missing information using graph-based completion~\cite{sengupta2020mm}, transformer aggregation~\cite{kini2024transhupr}, or diffusion-based fine-tuning on the human body ~\cite{fan2024diffusion}. However, despite these carefully designed models, evaluations in \cite{zhu2025probradarm3f, engel2025advanced, rahman2024mmvr} showed that subsequent models still cannot recover enough discarded features for fine-grained human posture modeling. This illustrates that PC's CFAR-based thresholding, while improving noise suppression, often leads to loss of subtle human reflections, resulting in PC's weak separation capability for non-human noise.

Both denoising strategies face a structural trade-off rooted in their input paradigms.  
Heatmap-based models retain complete motion information but embed overwhelming noise, which is statistically learned away.  
PC-based models explicitly suppress noise but sacrifice fine-grained human reflections essential for accurate skeletal reconstruction.  
This tension between information retention and noise suppression forms a persistent bottleneck in mmWave-based HPE.  

Our proposed \textbf{Person Parametric Physics-informed Representation (PPPR)} is designed to break this trade-off.  
By modeling each human joint as a Gaussian distribution parameterized by kinematic and electromagnetic properties, PPPR explicitly separates human-related signals from non-human noise.  
Its dual physics-informed objectives ensure that retained features obey biomechanical and radar-physical consistency, enabling maximal preservation of human motion information while minimizing both environmental and radar-induced noise.

\begin{table*}[h]
\centering
\label{tab:dataset_list}
\begin{adjustbox}{width=0.5\textwidth,center}
\setlength{\tabcolsep}{3pt}
\begin{tabular}{cccccc}
\hline
\textbf{Dataset} & \textbf{Year} & \textbf{View} & \textbf{Data Type} & \textbf{Frames} & \textbf{Full Public} \\ \hline
mm-Pose \cite{sengupta2022mmpose} & 2020 & Multi & Point Cloud & 40K & \faTimes \\
mmMesh \cite{xue2021mmmesh} & 2021 & Single & Point Cloud & 480K & \faTimes \\
mRI \cite{an2022mri} & 2022 & Single & Point Cloud & 160K & \faCheck \\
mmBody \cite{chen2022mmbody} & 2022 & Single & Point Cloud & 200K & \faCheck \\
MM-Fi \cite{yang2023mm} & 2023 & Single & Point Cloud & 320K & \faCheck \\

\cdashline{1-6}
RF-Pose \cite{yu2023rfpose} & 2018 & Multi & Heatmap & - & \faTimes \\
HuPR \cite{lee2023hupr} & 2023 & Multi & Heatmap & 141K & \faCheck \\
HIBER \cite{wu2022rfmasksimplebaselinehuman} & 2023 & Multi & Heatmap & 179K & \faTimes \\
MMVR \cite{rahman2024mmvr} & 2024 & Multi & Heatmap & 345K & \faCheck \\
XRF55 \cite{wang2024xrf55} & 2024 & Multi & Heatmap & 42.9K & \faCheck \\ \hline
\end{tabular}
\end{adjustbox}
\caption{Representative mmWave-based HPE datasets categorized by data modality.}
\end{table*}

\begin{table}[h]
\centering
\label{tab:input_paradigm_survey}
\begin{adjustbox}{width=0.5\textwidth,center}
\setlength{\tabcolsep}{3pt}
\begin{tabular}{lccc}
\hline
\textbf{Method} & \textbf{Year} & \textbf{Input Paradigm} & \textbf{Full OpenSource} \\ \hline
RF-Pose \cite{zhao2018through} & 2018&Heatmap&Yes\\
mm-Pose \cite{sengupta2020mm} & 2019 & PC & Yes \\
mmPose-NLP \cite{sengupta2022mmpose} & 2021 & PC &Yes \\
Cui et al. \cite{cui2021real} & 2021 & Other input & No \\
HuprModel \cite{lee2023hupr} & 2022 & Heatmap & Yes\\
mmGPE \cite{xue2023towards} & 2023 & Heatmap & No \\
RETR \cite{yataka2024retr} & 2024 & Heatmap & Yes \\
mmDiff \cite{fan2024diffusion} & 2024 & PC & Yes \\
TransHuPR \cite{kini2024transhupr} & 2024 & PC & No \\
ProbRadarM3F \cite{zhu2025probradarm3f} & 2025 & Heatmap & No \\
mmHPE \cite{wu2024mmhpe} & 2025 & Heatmap & No \\
MVDoppler-Pose \cite{choi2025mvdoppler} & 2025 & Other input & Yes \\
\hline
\end{tabular}
\end{adjustbox}
\caption{Survey of mmWave-based HPE systems by input paradigm.}
\end{table}

\subsection{Parametric Denoising in Computer Vision}
In the field of computer vision, parametric modeling has become a reliable approach for representing visual and spatial information in a compact and interpretable manner.
Parametric representations describe complex phenomena through compact parameter sets that separate intrinsic target properties from extrinsic environmental interference~\cite{Wu_2024_CVPR,xiang2017posecnn,wang2019densefusion}.  
This paradigm enables explicit denoising by design: instead of suppressing noise statistically, the representation itself encodes only the physical attributes relevant to the sensing target, allowing non-target variations to be naturally filtered during optimization.  
Such parameterization-based denoising has proven effective across several sensing modalities.
\subsubsection{Parameterization methods in Computer Vision}

\textbf{Dynamic scenes.}  
Gaussian Splatting (GS)~\cite{Wu_2024_CVPR} models visual scenes using Gaussian primitives that encode geometry and appearance parameters.  
By explicitly representing the spatial structure of a scene, GS can reconstruct clean images across diverse illumination and exposure conditions without requiring post-hoc filtering.  
Noise from lighting variations or sensor gain fluctuations is implicitly removed because these effects are not encoded in the geometric parameters.  
A differentiable rendering process enforces photometric consistency between observed images and synthesized projections, ensuring that the optimized parameters represent true geometric structure rather than noise-induced artifacts.

\textbf{Object pose.}  
In object-level perception, 6D pose estimation methods~\cite{xiang2017posecnn,wang2019densefusion} employ parametric formulations that describe an object's state using its translation and rotation.  
This design directly filters dense, noisy pixel data into concise geometric transformations, effectively isolating the object's spatial configuration from cluttered backgrounds, occlusions, or texture noise.  
By constraining the parameter space to physically meaningful transformations, these approaches suppress irrelevant environmental information and achieve denoising through structural abstraction.

\subsubsection{Limitation and Motivation.}
~

Although parametric denoising has achieved notable success in optical sensing, directly transplanting these paradigms to mmWave-based HPE is nontrivial.  
GS enforces only photometric consistency, which lacks constraints on the physical validity of electromagnetic propagation \cite{yang2025iradar}.  
As a result, its optimized parameters may yield visually coherent reconstructions that nevertheless violate radar signal physics—an especially critical issue under the coarse spatial resolution of mmWave sensing.  
Similarly, 6D pose methods rely solely on kinematic constraints, which are insufficient for disambiguating human joint reflections from dense clutter, as mmWave reflections from human limbs and surrounding structures often exhibit similar electromagnetic signatures.

To overcome these modality-specific limitations, we introduce the \textit{Person Parametric Physics-informed Representation (PPPR)}, a radar-oriented parametric input for mmWave-based HPE.
PPPR extends Gaussian parameterization with dual physics-informed constraints:
(1) \textit{Kinematic objectives} enforce biomechanical consistency to suppress spatial noise;
(2) \textit{Electromagnetic objectives} ensure radar-physical coherence by constraining Doppler–velocity relations and antenna array phase responses.
Coupled with a differentiable radar simulation, these constraints enable physically grounded denoising—preserving the complete human signal while mitigating environment- and radar-induced artifacts.

\subsection{Simulation- and Generation-based Methods}

\rev{Recent work explores improving cross-dataset generalization by synthesizing radar observations from other modalities (e.g., video-to-radar synthesis \cite{ahuja2021vid2doppler}) or by simulation-based data generation coupled with generative models (e.g., simulation + diffusion for mmWave data synthesis \cite{chen2023rf}). Broadly, these approaches aim to expand the effective training distribution by providing additional synthesized radar samples. In these pipelines, pose understanding is typically obtained by training downstream networks on the synthesized data using task losses. At the same time, the synthesis process is optimized toward its own reconstruction/generation objective.}

\rev{Complementary to data synthesis, PPPR focuses on a physics-informed intermediate representation that is explicitly parameterized by human structure and refined by kinematic constraints and electromagnetic consistency against observed Heatmaps. From a modeling perspective, PPPR targets \emph{per-sample} input formation/denoising via differentiable forward consistency, and can be used alongside synthesis-based pipelines as an input interface that enforces physical plausibility on radar observations.}

\section{Methodology}

\begin{figure*}[h]
  \centering
  \includegraphics[width=0.85\linewidth]{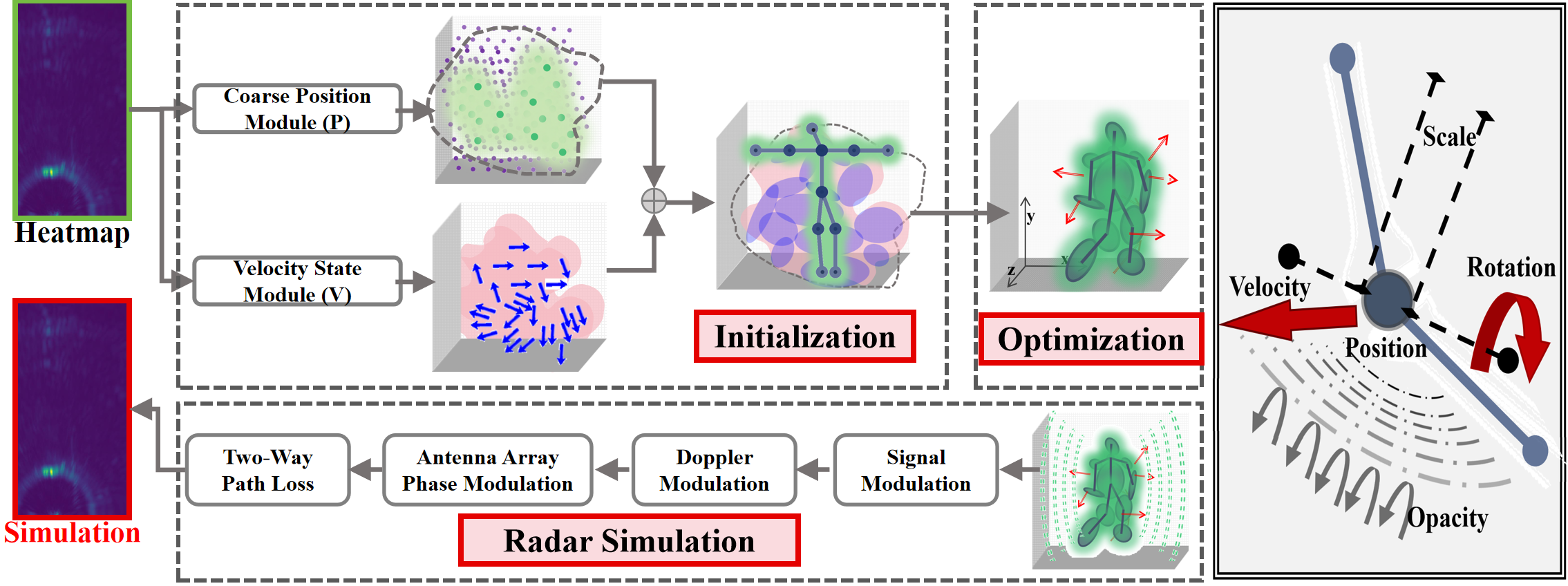}
  \caption{\rev{MmWave Human Parameterization (MHP), for a single person. Core components: Initialization (Sec.~\ref{sec:initialization}), Radar Simulation (Sec.~\ref{sec:radar_sim}), and Optimization (Sec.~\ref{sec:optimization}). The multi-person extension is described in Sec.~\ref{sec:multiperson}.} }
  \label{main_pipeline_single} 
\end{figure*}

In this section, we introduce the Person Parametric Physics-informed Representation (PPPR), which addresses the signal-noise challenge in mmWave-based HPE. PPPR's preparation process-MmWave Human Parameterization (MHP), as illustrated in Fig~\ref{main_pipeline_single}, consists of three stages:
(1) Initialization: Coarse joint positions and velocities are extracted from Heatmaps and encoded as Gaussian primitive parameters.
(2) Radar Simulation: A synthetic Heatmap is reconstructed from PPPR's parameter set through differentiable electromagnetic modeling. 
(3) Dual-Constraint Optimization: PPPR's parameters are refined under both kinematic constraints, enforcing biomechanical plausibility, and electromagnetic constraints, enforcing signal consistency. 

To the large number of mathematical symbols appearing in this section and other sections, we have provided a table in the Appendix summarizing the usage and physical meaning.

PPPR's parametric design also supports radar-agnostic deployment through FFT adaptation (Sec 4.2) and naturally extends to multi-person scenarios by incorporating inter-person collision constraints (Sec. 4.4).

\subsection{Initialization}
\label{sec:initialization}

Optimizing PPPR's parameter requires careful initialization to avoid local minima. Gaussian Splatting \cite{Wu_2024_CVPR} initializes geometric primitives from Structure-from-Motion (SfM) point clouds, leveraging multi-view consistency to obtain dense 3D geometry. However, mmWave Heatmaps lack such geometric constraints: sparse human reflections are embedded in environmental clutter, and single-view sensing cannot exploit multi-view consistency. Moreover, radar signals encode motion through Doppler shifts, requiring velocity initialization absent in visual reconstruction. We therefore design a physics-informed initialization that jointly extracts spatial positions and motion velocities, establishing motion-consistent starting states for subsequent optimization.

\subsubsection{Extracting Position and Velocity}

\noindent\textbf{Position extraction ($P$).} Given the radar Heatmap $H \in \mathbb{R}^{N_r \times N_a}$, where $N_r$ and $N_a$ denote the number of range and azimuth bins, we detect local maxima to identify candidate human reflections. Heatmaps obtained from common radars sometimes contain a very small number of NaN values, so following standard peak detection in radar processing \cite{hsieh2024multiperson,wu2020mmtrack}, low-pass filters are used to fill NaNs and retain peaks above an adaptive percentile threshold, yielding candidate peak set $\mathcal{P}_{\mathrm{cand}}$. Each peak at bin indices $(n_r, n_a)$ is 
converted to range $r$ and azimuth angle $\theta_{\mathrm{az}}$ through 
direct bin-to-physical mapping\footnote{In implementation, $r = (n_r/N_r) 
r_{\max}$ where $r_{\max}$ is the maximum detection range, and 
$\theta_{\mathrm{az}} = ((n_a/N_a) - 0.5)\text{FOV}_h$ where $\text{FOV}_h$ 
is the horizontal field-of-view. The factor 0.5 centers the coordinate system.}, 
then projected to Cartesian coordinates:
\begin{equation}
\small
(x, y, z) = (r\cos\theta_{\mathrm{az}}, r\sin\theta_{\mathrm{az}}, r\cos\theta_{\mathrm{el}}\sin\theta_{\mathrm{az}}),
\end{equation}
where for datasets with elevation information, $\theta_{\mathrm{el}}$ is computed 
from elevation bins, while for single-elevation configurations (common in our 
target datasets), $z$ is set to 0 during initialization due to dataset limitations.

For radars with elevation, $H \in \mathbb{R}^{N_r \times N_a \times N_e}$ where $N_e$ represents elevation bins. In this case, each peak at $(n_r, n_a, n_e)$ is converted to spherical coordinates $(r, \theta_{\mathrm{az}}, \theta_{\mathrm{el}})$ and projected to Cartesian coordinates via: $(x, y, z) = (r \cos\theta_{\mathrm{el}} \cos\theta_{\mathrm{az}}, r \cos\theta_{\mathrm{el}} \sin\theta_{\mathrm{az}}, r \sin\theta_{\mathrm{el}})$, where $\theta_{\mathrm{el}}$ is computed analogously to $\theta_{\mathrm{az}}$ from the elevation bin index. \cite{lee2023hupr,rahman2024mmvr}.

\textbf{Velocity extraction $V$.} Heatmap encodes motion through Doppler frequency shifts (Sec. 2.1). For each candidate peak, we extract the corresponding Doppler bin index $n_d$ and convert it to radial velocity following standard FMCW radar theory \cite{iovescu2020fundamentals}:
\begin{equation}
\small
v_r = \frac{\lambda \cdot n_d \cdot \text{PRF}}{2N_d},
\end{equation}
where $\lambda$ is the wavelength, pulse repetition frequency (PRF) represents the number of chirps transmitted per second, and $N_d$ is the number of Doppler bins. The radial velocity $v_r$ is decomposed into Cartesian components using the local angles: $(v_x, v_y, v_z) = v_r \cdot (\cos\theta_{\mathrm{el}} \cos\theta_{\mathrm{az}}, \cos\theta_{\mathrm{el}} \sin\theta_{\mathrm{az}}, \sin\theta_{\mathrm{el}})$, where $\theta_{\mathrm{el}}=0$ for planar configurations\footnote{For datasets without explicit Doppler dimensions \cite{wang2024xrf55}, we approximate velocity from spatial intensity gradients inspired by optical flow principles \cite{alfarano2024estimating}: $\hat{v}_r = \gamma \|\nabla H\|$, where $\gamma$ is a scaling factor calibrated to typical human motion speeds. This provides coarse velocity priors that are refined under electromagnetic reconstruction constraints during optimization (Sec. \ref{sec:radar_sim}). Implementation details are provided in Appendix A.}. This joint position-velocity extraction provides motion-consistent initialization\cite{wu2020mmtrack}.

\subsubsection{Parameterization}

To establish correspondence between extracted $(x, y, z)\&(v_x, v_y, v_z)$ and anatomical joints, we apply hierarchical clustering: peaks are grouped by spatial proximity and mapped to the human skeletal topology \cite{weng2022humannerf}, anchoring initialization near biomechanically plausible configurations. Each of the $N_j$ joints is then parameterized as a Gaussian primitive with parameter set $\Theta_j=\{\mathbf{p}_j,\mathbf{s}_j,\mathbf{q}_j,\mathbf{v}_j,\beta_j,\boldsymbol{\omega}_j\}$, categorized into three functional groups. This parametric representation enables compact, differentiable encoding while suppressing non-human noise during optimization (Limitation and Motivation in Sec 3.2).

\textbf{Geometric parameters} ($\mathbf{p}_j, \mathbf{s}_j, \mathbf{q}_j$) inherit vision-based modeling's spatial encoding \cite{Wu_2024_CVPR, xiang2017posecnn,wang2019densefusion} while adapting to skeletal constraints. \textit{Position} $\mathbf{p}_j\in\mathbb{R}^3$ defines the joint centroid. \textit{Scale} $\mathbf{s}_j\in\mathbb{R}^3$ captures anisotropic extent along three principal axes, forming diagonal matrix $\mathbf{S}_j=\mathrm{diag}(\mathbf{s}_j)$. Anisotropic modeling is essential for radar: elongated limb structures exhibit directional scattering patterns, unlike visual rendering, where isotropic Gaussians suffice \cite{iovescu2020fundamentals}. 
\textit{Rotation} $\mathbf{q}_j\in\mathbb{R}^4$ encodes 3D orientation using a quaternion representation, which is converted to a $3\times3$ rotation matrix $\mathbf{R}_j$ via standard formula\footnote{$\mathbf{R}_j = \mathbf{I} + 2w[\mathbf{u}]_{\times} + 2[\mathbf{u}]_{\times}^2$, where $\mathbf{q}_j=[w,\mathbf{u}]$ with scalar part $w$ and vector part $\mathbf{u}$ \cite{sola2017quaternion}.}. Compared to Euler angles, quaternions avoid numerical instabilities that occur when rotation axes align, enabling stable gradient-based optimization even when joints undergo large rotations during dynamic motions \cite{Wu_2024_CVPR}.

\textbf{Motion parameters} ($\mathbf{v}_j$) extend static visual reconstruction to dynamic radar scenarios. \textit{Velocity} $\mathbf{v}_j\in\mathbb{R}^3$ encodes instantaneous motion along Cartesian axes, initialized from Doppler extraction above. This parameter drives Doppler shift modeling in radar reconstruction (Sec. \ref{sec:radar_sim}), enabling velocity-dependent supervision absent in vision-based object modeling.

\textbf{Electromagnetic parameters} ($\beta_j, \boldsymbol{\omega}_j$) capture radar-specific signatures. \textit{Opacity} $\beta_j\in\mathbb{R}$ models the rate at which the influence of one joint on its surroundings decreases with distance from that joint. Unlike visual opacity, this represents scattering efficiency: larger body parts exhibit stronger returns than extremities \cite{iovescu2020fundamentals}. \textit{Doppler features} $\boldsymbol{\omega}_j\in\mathbb{R}^{N_d}$ encode frequency-domain characteristics across Doppler bins, enabling discrimination of spatially overlapping but kinematically distinct joints.

\textbf{Unified representation.} The spatial distribution $G$ of joint $j$ follows a 3D Gaussian \cite{Wu_2024_CVPR} with covariance $\boldsymbol{\Sigma}_j=\mathbf{R}_j\mathbf{S}_j\mathbf{S}_j^{\top}\mathbf{R}_j^{\top}$ and density:
\begin{equation}
\small
G_j(\mathbf{x})=(2\pi)^{-3/2}|\boldsymbol{\Sigma}_j|^{-1/2}\exp\!\left(-\tfrac{1}{2}(\mathbf{x}-\mathbf{p}_j)^{\top}\boldsymbol{\Sigma}_j^{-1}(\mathbf{x}-\mathbf{p}_j)\right),
\end{equation}
where the normalization $(2\pi)^{-3/2}$ ensures unit integral over 3D space. Each joint contributes a complex-valued radar return $\mathcal{R}$ based on radar processing standards \cite{iovescu2020fundamentals}:
\begin{equation}
\small
\mathcal{R}_j(\mathbf{x})=\alpha_j\,G_j(\mathbf{x})\cdot \exp\!\left(\mathbf{i}\,\boldsymbol{\omega}_j^{\top}\left[\mathbf{v}_j\odot(\mathbf{x}-\mathbf{p}_j)\right]\right),
\end{equation}
where $\mathbf{i}=\sqrt{-1}$, $\odot$ denotes element-wise multiplication,  $\alpha_j=\sigma(\beta_j)$, encoding normalized radar cross-section (RCS), and the exponential term encodes Doppler-induced phase shifts. This formulation preserves biomechanical structure through $(\mathbf{p}_j,\mathbf{s}_j,\mathbf{q}_j)$, electromagnetic observables through $(\alpha_j,\boldsymbol{\omega}_j)$, and motion dynamics through $\mathbf{v}_j$, establishing a unified physics-informed representation bridging kinematic and radar domains.

\subsection{Radar Pipeline Simulation}
\label{sec:radar_sim}

Biomechanical constraints alone (e.g., bone lengths, joint angles) cannot ensure that optimized PPPR parameters correspond to physically plausible radar returns. A kinematically valid pose may produce false radar signatures due to overfitting solely on prior kinematic knowledge. To address this, we design a differentiable radar simulation to enforce electromagnetic consistency. By reconstructing a synthetic Heatmap $H_{\mathrm{sim}}$ from PPPR parameters and performing differential analysis on the $H_{\mathrm{sim}}$ and the original Heatmap $H_{\mathrm{ori}}$, MHP can guide the optimized PPPR parameters to achieve electromagnetic consistency simultaneously.

\subsubsection{Signal reconstruction model.} 
Given optimized parameters $\{\Theta_j\}_{j=1}^{N_j}$ for $N_j$ joints, the simulated $H_{\mathrm{sim}}$ aggregates per-joint complex-valued radar returns. Each joint's contribution is modulated by four physically grounded operators that capture amplitude attenuation and phase shifts in radar signal propagation.
\begin{equation}
\small
H_{\mathrm{sim}}=\sum_{j=1}^{N_j} M_{\mathrm{atten}}^{(j)}\; M_{\mathrm{range}}^{(j)}\; M_{\mathrm{Dopp}}^{(j)}\; M_{\mathrm{angle}}^{(j)}\; \mathcal{R}_j,
\end{equation}
where $\mathcal{R}_j$ is the joint-wise complex radar return defined in Sec.~\ref{sec:initialization}, and each modulation operator $M_{\cdot}^{(j)}$ encodes a specific aspect of electromagnetic propagation. We describe these operators sequentially, beginning with amplitude attenuation, followed by three phase shift components:

\begin{itemize}
\item \textbf{Two-way path attenuation} $M_{\mathrm{atten}}^{(j)}$ models 
amplitude decay due to round-trip propagation. Following radar equation 
principles \cite{piersanti2012millimeter}, received 
power is proportional to the inverse fourth power of range $d_j = \|\mathbf{p}_j\|$. We assume the radar is located at the coordinate origin. Joint positions $\mathbf{p}_j$ therefore directly represent range vectors from the radar:
\begin{equation}
\small
M_{\mathrm{atten}}^{(j)} \propto d_j^{-4},
\end{equation}
where the proportionality captures both outbound and inbound spherical spreading 
($d_j^{-2}$ each direction) and target scattering efficiency. In our implementation, 
we incorporate this attenuation through the joint-wise complex return $\mathcal{R}_j$, where the proportionality constant is absorbed into the opacity parameter 
$\alpha_j$, yielding the normalized attenuation factor $M_{\mathrm{atten}}^{(j)} = d_j^{-4} / d_{\max}^{-4}$ 
with $d_{\max}$ being the maximum detection range.

\item \textbf{Range-dependent phase shift} $M_{\mathrm{range}}^{(j)}$ models beat frequency generation in FMCW radar (Range FFT in Sec. 2.1). With round-trip delay $\tau_j=2d_j/c$ and chirp slope $S=B/T_c$ (bandwidth $B$, chirp duration $T_c$), the beat frequency is $f_{\mathrm{beat}}^{(j)}=S\tau_j$, yielding phase modulation \cite{richards2005fundamentals}:
\begin{equation}
\small
M_{\mathrm{range}}^{(j)}=\exp\!\left(\mathbf{i}\,2\pi f_{\mathrm{beat}}^{(j)}\,\tau_j\right),
\end{equation}
where $\mathbf{i}=\sqrt{-1}$. This links the joint position to the Range FFT output (Sec. 2.1).

\item \textbf{Doppler phase shift} $M_{\mathrm{Dopp}}^{(j)}$ captures velocity-induced frequency modulation. Radial velocity $v_{r,j}=\mathbf{v}_j\cdot(\mathbf{p}_j/d_j)$ induces Doppler shift $\Delta f_{\mathrm{Dopp}}^{(j)}=2v_{r,j}/\lambda$ (Doppler FFT in Sec. 2.1), producing phase modulation across the coherent processing interval of duration $T_{\mathrm{frame}}$ \cite{chen2006micro}:
\begin{equation}
\small
M_{\mathrm{Dopp}}^{(j)}=\exp\!\left(\mathbf{i}\,2\pi \Delta f_{\mathrm{Dopp}}^{(j)}\, T_{\mathrm{frame}}\right).
\end{equation}
Explicit velocity-dependent phase modulation prevents optimization from fabricating motion through geometric or opacity adjustments alone, enabling disambiguation of spatially overlapping joints with distinct kinematics.

\item \textbf{Antenna array phase shifts} $M_{\mathrm{angle}}^{(j)}$ simulate angular processing via element-wise phase differences (Angle FFT in Sec. 2.1). With azimuth and elevation angles computed from joint position as $\theta_{\mathrm{az},j}=\arctan(p_{j,y}/p_{j,x})$ and $\theta_{\mathrm{el},j}=\arctan(p_{j,z}/\sqrt{p_{j,x}^2+p_{j,y}^2})$, and antenna spacings $l_{\mathrm{az}}, l_{\mathrm{el}}$, the phase differences are \cite{van2002optimum}:
\begin{equation}
\small
\Delta\phi_{\mathrm{az},j}=\tfrac{2\pi}{\lambda}\,l_{\mathrm{az}}\sin(\theta_{\mathrm{az},j}),\quad
\Delta\phi_{\mathrm{el},j}=\tfrac{2\pi}{\lambda}\,l_{\mathrm{el}}\sin(\theta_{\mathrm{el},j}),
\end{equation}
yielding combined phase modulation:
\begin{equation}
\small
M_{\mathrm{angle}}^{(j)}=\exp\!\left(\mathbf{i}\,\Delta\phi_{\mathrm{az},j}\right)\exp\!\left(\mathbf{i}\,\Delta\phi_{\mathrm{el},j}\right).
\end{equation}
This reproduces azimuth/elevation selectivity in the Angle FFT (Sec. 2.1).
\end{itemize}

Together, these four operators decompose radar signal formation into amplitude decay ($M_{\mathrm{atten}}$) and three phase shift components ($M_{\mathrm{range}}, M_{\mathrm{Dopp}}, M_{\mathrm{angle}}$), bridging PPPR parameters to original Heatmap measurements through physically interpretable transformations.

\subsubsection{Radar-agnostic calibration.}
~

\begin{figure*}[h]
  \centering
  \includegraphics[width=0.8\linewidth]{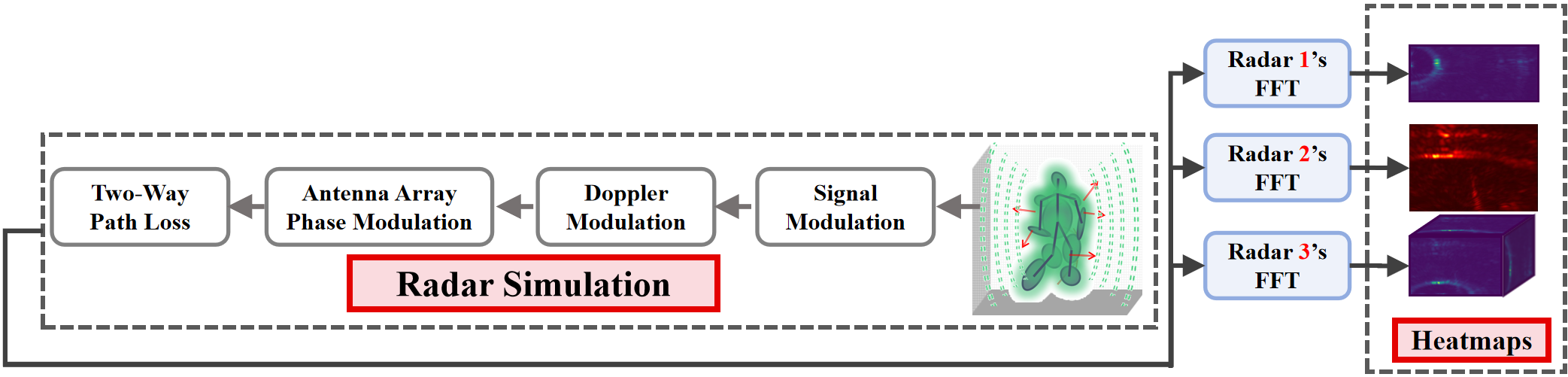}
  \caption{MHP's adaptability to different radar FFT processing used on datasets.}
  \label{FFT} 
\end{figure*}

The simulated complex returns represent raw signals before Fast Fourier Transform processing. Real-world radar systems employ different FFT configurations depending on radar design: different manufacturers use varying chirp counts, sampling rates, and windowing functions \cite{rahman2024mmvr,lee2023hupr,wang2024xrf55}. To enable cross-radar deployment, we apply a radar-specific calibration step as illustrated in Fig. \ref{FFT}: the FFT configuration, including chirp parameters and window functions, corresponding to the target radar, is applied to the simulated signals, producing $H_{\mathrm{sim}}$ with identical dimensionality and spectral characteristics as the original Heatmap $H_{\mathrm{ori}}$. This calibration is analogous to camera intrinsic calibration in computer vision—a necessary adaptation to radar specifications that preserves the underlying physical model. Consequently, electromagnetic reconstruction loss on $H_{\mathrm{sim}}$ and $H_{\mathrm{ori}}$ can be computed directly regardless of radar, enabling a unified PPPR model across different deployments without retraining.

\subsection{Optimization}
\label{sec:optimization}

PPPR parameters are optimized under dual physics-informed objectives that enforce both biomechanical plausibility and electromagnetic consistency. Without kinematic constraints, optimization may converge to poses that match radar observations but violate human anatomical structure (e.g., elongated limbs, impossible joint configurations). On the other hand, without electromagnetic constraints, optimization may produce kinematically valid but radar-inconsistent poses (e.g., limbs positioned at electromagnetically implausible angles). We therefore couple these complementary constraints to guide optimization toward solutions that satisfy both kinematic laws and radar signal physics.

\textbf{Kinematic constraints} regularize PPPR parameters through three biomechanical priors that enforce universal laws of human skeletal structure \cite{weng2022humannerf,jiang2022neuman}. Let $\mathcal{E}$ denote the set of bone edges connecting adjacent joints, and $\mathcal{A}$ denote the set of joint angle triplets $(m, n, o)$ where $n$ is the vertex joint connecting bones $mn$ and $no$. We define:

\begin{itemize}
\item \textit{Bone length consistency}: Human skeletal dimensions remain constant over short time scales. For each bone $(m,n) \in \mathcal{E}$ with reference length $\ell_{mn}$ (initialized from average in datasets \cite{rahman2024mmvr}), we penalize deviations:
\begin{equation}
\small
\mathcal{L}_{\mathrm{bone}}=\sum_{(m,n)\in \mathcal{E}}\!\big(\|\mathbf{p}_m-\mathbf{p}_n\|-\ell_{mn}\big).
\end{equation}

\item \textit{Rigid bone motion}: Bones do not stretch or compress during motion. The relative velocity component along bone direction $\hat{\mathbf{b}}_{mn}=(\mathbf{p}_m-\mathbf{p}_n)/\|\mathbf{p}_m-\mathbf{p}_n\|$ should be zero:
\begin{equation}
\small
\mathcal{L}_{\mathrm{rigid}}=\sum_{(m,n)\in \mathcal{E}}\!\big\|(\mathbf{v}_m-\mathbf{v}_n)\cdot\hat{\mathbf{b}}_{mn}\big\|.
\end{equation}
This permits rotation around joints while preventing non-physical bone deformation.

\rev{The $\mathcal{L}_{\mathrm{bone}}$ anchors bone lengths to reference values, while $\mathcal{L}_{\mathrm{rigid}}$ enforces rigid-bone motion by penalizing relative velocity along the bone direction (i.e., adjacent joints should have consistent velocity components along $\hat{\mathbf{b}}_{mn}$), thereby avoiding non-physical bone stretching/compression during motion.}

\item \textit{Joint angle limits}: Articulation is constrained within physiologically feasible ranges \cite{akhter2015pose}. For joint angle triplet $(m,n,o) \in \mathcal{A}$, the angle $\theta_{mno}$ between bones $\mathbf{p}_m-\mathbf{p}_n$ and $\mathbf{p}_o-\mathbf{p}_n$ is restricted to $[0, \theta_{\max}]$:
\begin{equation}
\small
\mathcal{L}_{\mathrm{joint}}=\sum_{(m,n,o)\in \mathcal{A}}\!\max(0,|\theta_{mno}|-\theta_{\max})^2.
\end{equation}
\end{itemize}

The kinematic objective aggregates these three constraints:
\begin{equation}
\small
\mathcal{L}_{\mathrm{kine}}=\mathcal{L}_{\mathrm{bone}}+\mathcal{L}_{\mathrm{rigid}}+\mathcal{L}_{\mathrm{joint}}.
\end{equation}

\textbf{Electromagnetic constraint} enforces that optimized PPPR parameters reproduce radar signatures. The simulated Heatmap $H_{\mathrm{sim}}$ (Sec. \ref{sec:radar_sim}) is compared against the original Heatmap $H_{\mathrm{ori}}$  using an IoU-based reconstruction loss. Standard losses like MSE are suboptimal for radar Heatmaps, which exhibit extreme sparsity: over 95\% of spatial cells contain only background noise due to limited human body coverage in the sensing volume. These loss treats all locations equally, causing optimization to fit empty regions rather than informative human reflections. We therefore adopt IoU loss \cite{cheng2021boundary}, which emphasizes overlap between high-intensity regions:
\begin{equation}
\small
\mathcal{L}_{\mathrm{EM}}=1-\frac{|\mathcal{B}_{\mathrm{sim}}\cap \mathcal{B}_{\mathrm{ori}}|}{|\mathcal{B}_{\mathrm{sim}}\cup \mathcal{B}_{\mathrm{ori}}|},
\label{eq:em_iou}
\end{equation}
where $\mathcal{B}_{\mathrm{sim}}$ and $\mathcal{B}_{\mathrm{ori}}$ are binary masked Heatmaps extracted by adaptive thresholding from $\mathcal{H}_{\mathrm{sim}}$ and $\mathcal{H}_{\mathrm{ori}}$: cells with intensities above the top $\tau_{\mathrm{pct}}$-th percentile are marked as foreground\footnote{Binary masks are computed as $\mathcal{B} = \{(r,a) \mid H(r,a) > \text{percentile}(H, 100-\tau_{\mathrm{pct}})\}$, where $(r,a)$ denotes range-azimuth indices. In addition to the ablation experiment, $\tau_{\mathrm{pct}}$ was set to 10\%.}. This IoU formulation captures spatial alignment of radar returns while remaining robust to intensity scale variations.

\textbf{Unified optimization} balances kinematic and electromagnetic objectives through a weighted combination:
\begin{equation}
\small
\mathcal{L}_{\mathrm{total}}=w_{\mathrm{EM}}\mathcal{L}_{\mathrm{EM}}+w_{\mathrm{kine}}\mathcal{L}_{\mathrm{kine}},
\label{eq:total_loss}
\end{equation}
where $w_{\mathrm{EM}}$ and $w_{\mathrm{kine}}$ are hyperparameters controlling the trade-off between electromagnetic fidelity and kinematic plausibility\footnote{In addition to the ablation experiment, we set $w_{\mathrm{EM}}=0.5$ and $w_{\mathrm{kine}}=0.5$ across all experiments. These weights are not constrained to sum to unity.}. Parameters $\{\Theta_j\}_{j=1}^{N_j}$ are optimized via gradient descent . This dual-constraint framework ensures optimized poses satisfy both human biomechanics and radar propagation physics, yielding PPPR representations that encode clean skeletal features while suppressing environmental and radar hardware noise. The framework naturally extends to multi-person scenarios (Sec. \ref{sec:multiperson}) by incorporating inter-person collision constraints.

\subsection{Multi-Person Extension}
\label{sec:multiperson}

\begin{figure*}[h]
  \centering
  \includegraphics[width=0.8\linewidth]{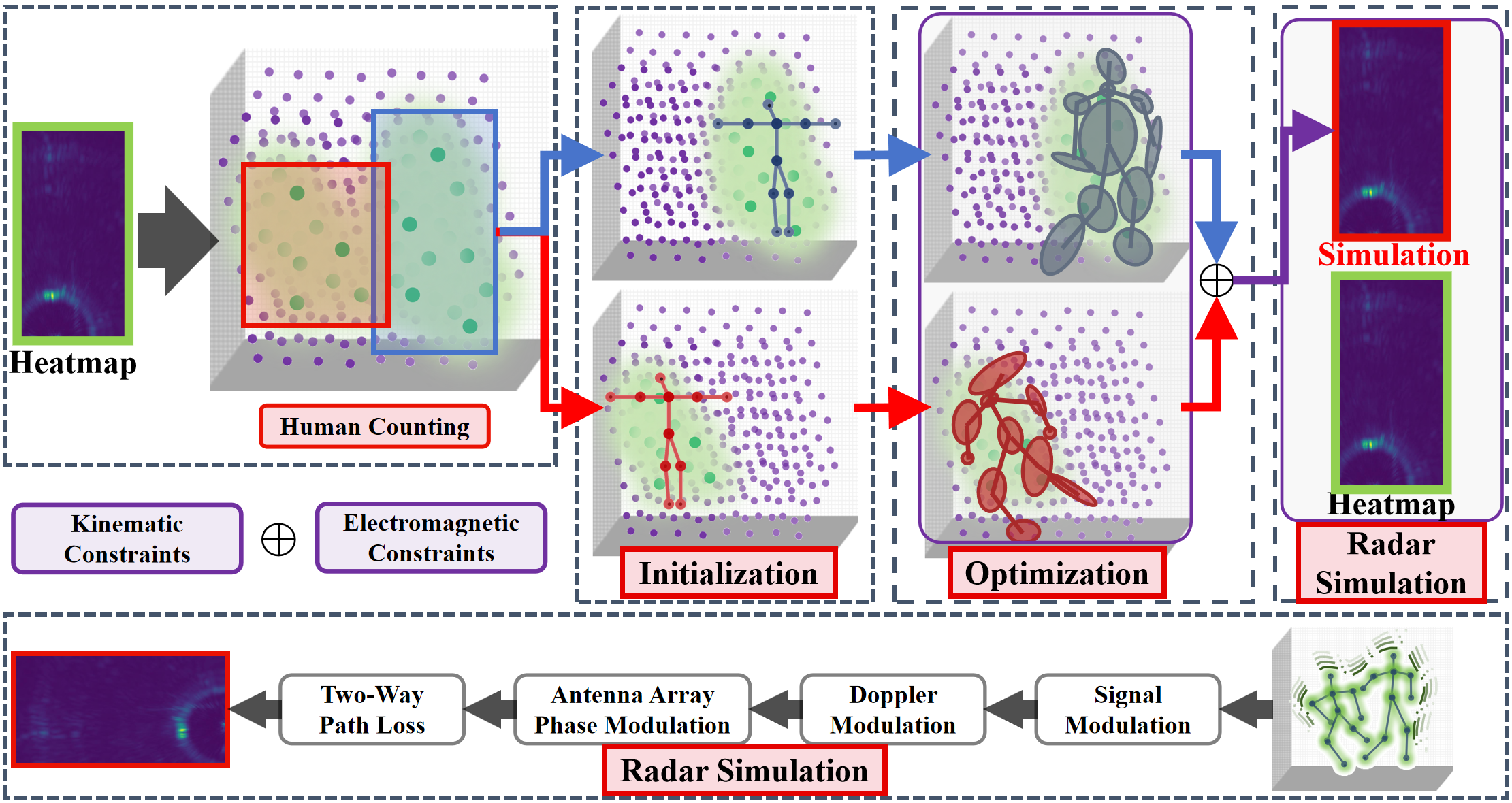}
  \caption{MmWave Human Parameterization (MHP) for multi-person scenarios. Two green-labeled Heatmaps are identical, both are the original input $H_{\mathrm{ori}}$. Key extensions include person counting to determine instance number and inter-person collision constraints to prevent skeletal overlap.} 
  \label{main_pipeline_multi} 
\end{figure*}

Multi-person scenarios introduce two challenges beyond single-person HPE: (1) determining the number of individuals present to initialize the correct number of PPPR instances, and (2) preventing spatial overlap between estimated skeletons during optimization. Radar reflections from multiple humans create overlapping signal patterns in the Heatmap, complicating both detection and association \cite{hsieh2024multiperson}. We address these challenges through two extensions to the PPPR framework: person counting via clustering-based detection, and inter-person collision constraints that enforce spatial separation during optimization (Fig. \ref{main_pipeline_multi}).

\textbf{Person counting} determines the number of PPPR instances $N_{\mathrm{person}}$ to initialize from the original Heatmap. We adopt a two-stage approach: ETCM-CFAR \cite{yang2025spatial} performs temporal-correlation-based peak detection to identify candidate human reflections, followed by DBSCAN \cite{ester1996density} spatial clustering to group peaks belonging to the same individual. A lightweight MLP classifier refines the discrete person count from clustered features\footnote{The MLP takes cluster statistics (centroid positions, spatial extent, mean intensity) as input and outputs person count. This module is trained on multi-person Heatmap samples with ground-truth counts, and has room for optimization, but it is not the main focus of this paper; it is only used as a functional module in multi-person scenarios. Architecture details are in the Appendix.}. This provides the instance count $N_{\mathrm{person}}$ for subsequent PPPR initialization. This module serves as a functional component enabling multi-person validation.

\textbf{Inter-person constraints} prevent skeletal overlap when multiple PPPR instances are optimized jointly. Let $s \in \{1, \ldots, N_{\mathrm{person}}\}$ index individuals, with skeleton centroids $\mathbf{c}_s = \frac{1}{N_j}\sum_{n=1}^{N_j} \mathbf{p}_{s,n}$ (average over $N_j$ joint positions $\mathbf{p}_{s,n}$). We enforce two spatial separation penalties:

\begin{itemize}
\item \textit{Centroid separation} prevents individuals from occupying the same spatial region. For distinct persons $s \neq t$, we penalize center proximity:
\begin{equation}
\small
\mathcal{L}_{\mathrm{sep}}=\sum_{s \neq t}\max\!\left(0, d_{\mathrm{sep}} - \|\mathbf{c}_s-\mathbf{c}_t\|\right),
\end{equation}
where $d_{\mathrm{sep}}$ is the minimum allowed centroid distance (typically 0.5m, reflecting typical human body width).

\item \textit{Joint-level collision avoidance} prevents limb overlap between different individuals. For all joint pairs $(m,n)$ across distinct persons $(s,t)$, we penalize close proximity:
\begin{equation}
\small
\mathcal{L}_{\mathrm{coll}}=\sum_{s \neq t}\sum_{m,n \in \{1,\ldots,N_j\}}\max\!\left(0, d_{\mathrm{joint}} - \|\mathbf{p}_{s,m}-\mathbf{p}_{t,n}\|\right),
\end{equation}
where $d_{\mathrm{joint}}$ is the minimum allowed joint distance (typically 0.1m)\footnote{These thresholds are conservative estimates based on human body dimensions. The squared term outside $\max(\cdot)$ provides smooth gradients when constraints are active.}.
\end{itemize}

The multi-person kinematic objective extends the single-person kinematic objective in Sec.~\ref{sec:optimization} by aggregating constraints across all individuals and adding inter-person penalties:
\begin{equation}
\small
\mathcal{L}_{\mathrm{kine,multi}}=\sum_{s=1}^{N_{\mathrm{person}}}\!\left(\mathcal{L}_{\mathrm{bone},s}+\mathcal{L}_{\mathrm{rigid},s}+\mathcal{L}_{\mathrm{joint},s}\right)+\mathcal{L}_{\mathrm{sep}}+\mathcal{L}_{\mathrm{coll}},
\end{equation}
where each per-person term $\mathcal{L}_{\cdot,s}$ follows the single-person definition but operates on person $s$'s parameters. Multi-person optimization then uses the unified objective with $\mathcal{L}_{\mathrm{kine,multi}}$ replacing $\mathcal{L}_{\mathrm{kine}}$. These constraints preserve individual identity through spatial separation: when two persons approach closely (e.g., $\|\mathbf{c}_s-\mathbf{c}_t\| < d_{\mathrm{sep}}$), $\mathcal{L}_{\mathrm{sep}}$ generates repulsive gradients that push skeletons apart, while $\mathcal{L}_{\mathrm{coll}}$ prevents limb-level interpenetration. This ensures distinct skeletal estimates even when radar reflections partially overlap in the Heatmap.

\section{Experiments}

\subsection{Experimental Setup}

\subsubsection{Datasets}

\begin{table}[h]
\centering
\begin{adjustbox}{width=0.9\textwidth,center}
\setlength{\tabcolsep}{1pt}
\begin{tabular}{ccccccccc}
\hline
\textbf{Dataset} & \textbf{Year} & \textbf{Data Type}  & \textbf{mmWave radar} & \textbf{Heatmap Shape} & \textbf{FFT Configuration} & \textbf{Subjects} & \textbf{Scenes} & \textbf{Frames} \\ \hline
XRF55 \cite{wang2024xrf55} & 2024 & Heatmap & TI IWR6843ISK & 256×128 & Doppler+Angle & 39 & 4 & 42.9K \\
HuPR \cite{lee2023hupr} & 2023 & Heatmap & TI IWR1843BOOST & 64×64×8 & 3D Unified & 6 & 1 & 14.1K \\
MMVR \cite{rahman2024mmvr} & 2024 & Heatmap & TI AWR2243 & 256×128 & 3D Unified & 25 & 6 & 345K \\ \hline
\end{tabular}
\end{adjustbox}
\caption{Evaluation datasets with Heatmap format specifications and radar configurations.}
\label{tab:test_dataset}
\end{table}
We evaluate PPPR on three publicly available mmWave-based HPE datasets that provide Heatmap inputs with 3D joint annotations: MMVR \cite{rahman2024mmvr}, XRF55 \cite{wang2024xrf55}, and HuPR \cite{lee2023hupr} (Table~\ref{tab:test_dataset}).

\textbf{MMVR} employs TI AWR2243 radar with 3D Unified FFT: Range FFT extracts distance, Doppler FFT extracts velocity, and Angle FFT extracts both azimuth and elevation. After accumulating and compressing the Doppler dimension, a 2D data map is produced. The dataset encompasses six distinct scenes: open-foreground scenarios in a 9.5×10m conference room with minimal furniture, and cluttered multi-room environments including a lobby, two small offices, and two conference rooms.

\textbf{XRF55} uses TI IWR6843ISK radar with Doppler+Angle FFT: Range FFT and Doppler FFT extract range-velocity information, while Angle FFT processes only azimuth, producing a 2D range-azimuth representation. The dataset collects data across four indoor scenes with consistent radar placement but varying environmental layouts: Scene 1 features office settings with partitioned workspaces, Scene 2 presents an open laboratory with glass walls, and Scenes 3-4 exhibit different furniture arrangements in conference-style spaces. 

\textbf{HuPR} employs TI IWR1843BOOST radar with 3D Unified FFT producing 64×64×8 range-azimuth-elevation Heatmap. It provides controlled single-room recordings with 6 subjects performing systematic pose variations.

\subsubsection{Evaluation Metrics}

Following established HPE evaluation protocols \cite{ionescu2013human3}, we measure pose estimation accuracy using: (1) Mean Absolute Joint Position Error (\textbf{MAJPE}), the average Euclidean distance between predicted and ground-truth joint positions across all joints and frames; (2) Procrustes-Aligned MAJPE (\textbf{PA-MAJPE} ), which neutralizes global pose transformations to focus on skeletal structure accuracy. Lower values indicate better performance for both metrics.

\subsubsection{Tested HPE models}

We evaluate PPPR by comparing against conventional input paradigms-Heatmap and PC-across four representative HPE architectures:

\begin{itemize}
\item \textbf{RETR} \cite{yataka2024retr}: Transformer-based architecture with multi-head self-attention, operating on Heatmap inputs.
\item \textbf{mmDiff} \cite{fan2024diffusion}: Diffusion model with iterative denoising, operating on PC inputs derived via CFAR.
\item \textbf{HuprModel} \cite{lee2023hupr}: Multi-scale CNN with hierarchical feature pyramids, operating on Heatmap inputs.
\item \textbf{PoseformerV2} \cite{zhao2023poseformerv2}: Temporal transformer originally designed for vision-based HPE, adapted to mmWave via learned feature transformation from Heatmap inputs.
\end{itemize}

All HPE model architectures remain unchanged across input paradigms to isolate the contribution of input design. Training uses identical hyperparameters, data splits, and optimization strategies for fair comparison.

\subsubsection{Experimental Protocols}

Table~\ref{tab:train_test_splits} summarizes our evaluation protocols across five configurations: within-scene performance, cross-scene environment robustness, cross-dataset radar transferability, and multi-person extensions. Each configuration isolates specific PPPR capabilities: within-scene establishes accuracy, cross-scene evaluates environment-decoupling, cross-dataset validates radar-agnostic calibration, and multi-person tests instance separation.

\begin{table}[h]
\centering
\begin{adjustbox}{width=0.85\textwidth,center}
\setlength{\tabcolsep}{1pt}
\begin{tabular}{lcccc}
\hline
\textbf{Evaluation Type} & \textbf{Train} & \textbf{Test} & \textbf{Purpose} & \textbf{Section} \\ \hline
Within-scene & All datasets\&scenes & Same datasets\&scene & Baseline performance & 5.2.1 \\
Cross-scene & MMVR Scene 1 & MMVR Scenes 2-5 & Environment denoise & 5.2.2 \\
Cross-dataset & MMVR & XRF55, HuPR & Radar denoise & 5.2.3 \\
Within-scene(Multi) & MMVR\&XRF55 Multi & Same Multi & Multi-person performance & 5.3.3 \\ 
Cross-scene(Multi)  & MMVR Multi Scene 1 & MMVR Multi Scenes 2-5 & Environment denoise & 5.3.4 \\ \hline
\end{tabular}
\end{adjustbox}
\caption{Train/test configurations for comprehensive PPPR evaluation across single-person and multi-person scenarios.}
\label{tab:train_test_splits}
\end{table}

\rev{We follow standard evaluation protocols in prior mmWave-based HPE works~\cite{fan2024diffusion,yataka2024retr,lee2023hupr}, with all train/test configurations summarized in Table~\ref{tab:train_test_splits}.}

\subsection{Single-Person HPE Scenarios}

We conduct a number of experiments to study the effectiveness of PPPR in different HPE scenarios: establishing performance within datasets (Sec. 5.2.1), assessing environment robustness via cross-scene evaluation (Sec. 5.2.2), and validating cross-radar transferability via cross-dataset evaluation (Sec. 5.2.3).

\subsubsection{Within-Dataset Performance}

\rev{To validate PPPR's core contribution as a physics-informed parametric intermediate representation, we evaluate conventional inputs (Heatmap, PC) versus PPPR across four HPE models on single-person scenarios. Additionally, we test PPPR-enhanced variants: PPPR-Heatmap (PPPR's parameters reprojected to Heatmap format) and PPPR-PC (PPPR-Heatmap reprojected to Point Cloud), which enable compatibility with the existing HPE baseline.}

\begin{table}[h]
\centering
\label{tab:PPPR_improvement}
\begin{adjustbox}{width=\textwidth,center}
\setlength{\tabcolsep}{2pt}
\begin{tabular}{ccccccccc}
\hline
& & & \multicolumn{2}{c}{\textbf{MMVR}} & \multicolumn{2}{c}{\textbf{HuPR}} & \multicolumn{2}{c}{\textbf{XRF55}} \\
\multirow{-2}{*}{\textbf{HPE Model}} & \multirow{-2}{*}{\textbf{Input}} & \multirow{-2}{*}{\textbf{Size}} & \textbf{MAJPE$\downarrow$} & \textbf{PA-MAJPE$\downarrow$} & \textbf{MAJPE$\downarrow$} & \textbf{PA-MAJPE$\downarrow$} & \textbf{MAJPE$\downarrow$} & \textbf{PA-MAJPE$\downarrow$} \\ \hline

& Heatmap & 76.9M & 73.05 & 66.97 & 78.09 & 72.54 & 81.77 & 78.90 \\
& PPPR & 12.9M & 64.71 (-8.34) & 58.91 (-8.06) & 73.74 (-4.35) & 67.70 (-4.84) & 72.00 (-9.77) & 68.71 (-10.19) \\
\multirow{-3}{*}{RETR \cite{yataka2024retr}} & PPPR-Heatmap & 76.9M & 69.30 (-3.75) & 61.60 (-5.37) & 75.90 (-2.19) & 70.84 (-1.70) & 77.51 (-4.26) & 71.90 (-7.00) \\
\cdashline{2-9}

& Heatmap  & 524.9M & 70.24 & 64.88 & 65.37 & 58.11 & 80.26 & 76.06 \\
& PPPR & 370.6M & \underline{60.34} (-9.90) & 55.07 (-9.81) & \underline{62.27} (-3.10) & \underline{53.64} (-4.47) & 71.04 (-9.22) & 67.83 (-8.23) \\
\multirow{-3}{*}{HuprModel \cite{lee2023hupr}} & PPPR-Heatmap & 524.9M & 63.56 (-6.68) & 57.80 (-7.08) & 63.21 (-2.16) & 55.57 (-2.54) & 75.72 (-4.54) & 69.30 (-6.76) \\
\cdashline{2-9}

& PC & 362.8M & 68.62 & 62.11 & 75.54 & 70.02 & 79.06 & 74.31 \\
& PPPR & 91.2M & 61.57 (-7.05) & \underline{54.99} (-7.12) & 70.95 (-4.59) & 62.44 (-7.58) & \underline{69.75} (-9.31) & \underline{67.03} (-7.28) \\
\multirow{-3}{*}{mmDiff \cite{fan2024diffusion}} & PPPR-PC & 87.2M & 67.90 (-0.72) & 59.07 (-3.04) & 73.98 (-1.56) & 69.11 (-0.91) & 70.47 (-8.59) & 69.17 (-5.14) \\
\cdashline{2-9}

PoseformerV2 \cite{zhao2023poseformerv2} & PPPR  & 112.0M & 62.42 & 56.93 & 63.34 & 54.87 & 72.07 & 69.15 \\ \hline
\end{tabular}
\end{adjustbox}
\caption{Single-person performance comparison. We compare PPPR, PPPR-enhanced Heatmap (PPPR-Heatmap), and PPPR-enhanced PC (PPPR-PC) against these models' original inputs (Heatmap, PC) using four HPE models across three datasets. Parentheses show MAJPE reduction.}
\end{table}

Table 5 presents single-person performance across four HPE models on three datasets. PPPR consistently achieves 4-10mm MAJPE reduction while using 29-83\% fewer parameters. Even when reprojected to conventional formats as PPPR-Heatmap or PPPR-PC, denoised representations maintain 2-7mm improvements over raw inputs (Heatmap and PC).

These results validate PPPR's core mechanism: parametric modeling under dual physics constraints effectively separates human skeletal features from noise sources. Consistent gains across diverse HPE architectures (transformer, CNN, diffusion) confirm that the advantage stems from input-level noise suppression. PPPR enables plug-and-play integration with existing HPE models without architectural modifications, providing immediate performance improvements with reduced computational overhead. This compatibility facilitates adoption in deployed systems through input preprocessing alone. It is worth mentioning that, based on PPPR, HPE models from the computer vision field (e.g., PoseformerV2) can also be applied with promising results, suggesting PPPR is a flexible and high signal-to-noise-ratio (SNR) solution for mmWave-based HPE.

\subsubsection{Cross-Scene Evaluation}

MMVR dataset is collected in 6 different scenes using the same radar, and we design experiments to evaluate PPPR's capacity for suppressing environment noises exhibited in different scenes. Specifically, we train three HPE models on MMVR Scene 1 and evaluate on Scenes 2-5 without fine-tuning. The last scenario had fewer samples and mostly involved multiple people participating simultaneously.

\begin{table}[h]
\centering
\setlength{\tabcolsep}{1pt}
\begin{adjustbox}{width=0.5\textwidth,center}
\begin{tabular}{lcccccc}
\hline
\multirow{2}{*}{\textbf{Test Scene}} & \multicolumn{2}{c}{\textbf{mmDiff}} & \multicolumn{2}{c}{\textbf{HuPRmodel}} & \multicolumn{2}{c}{\textbf{RETR}} \\ \cline{2-7}
& \textbf{PC} & \textbf{PPPR} & \textbf{Heatmap} & \textbf{PPPR} & \textbf{Heatmap} & \textbf{PPPR} \\ \hline
Scene 2  & 92.4 & 75.1 & 95.8 & 76.3 & 88.7 & 73.0 \\
Scene 3  & 93.1 & 75.8 & 96.2 & 77.0 & 89.0 & 74.2 \\
Scene 4  & 92.8 & 75.4 & 95.5 & 76.5 & 89.6 & 74.4 \\
Scene 5  & 93.6 & 75.6 & 96.0 & 76.8 & 89.2 & 73.8 \\ \hline
\textbf{Mean} & \textbf{93.0} & \textbf{75.5} & \textbf{95.9} & \textbf{76.7} & \textbf{89.1} & \textbf{73.9} \\ \hline
\end{tabular}
\end{adjustbox}
\caption{Performance comparison (MAJPE) between PPPR and conventional inputs via three HPE models in cross-scene scenarios. All models were trained on MMVR's Scene 1 and tested on MMVR's Scenes 2 to 5 without fine-tuning.}
\label{tab:cross_scene}
\end{table}

As can be seen from Table~\ref{tab:cross_scene}, traditional inputs such as PC or Heatmap degrade 22-27mm (30-36\% relative increase) when deployed to unseen scenes, while PPPR maintains 74-77mm MAJPE with minimal scene variance (<2mm standard deviation).

This stability confirms PPPR's environment-decoupling mechanism. 
Conventional inputs overfit to furniture arrangements and wall reflection characteristics during training, causing performance collapse when encountering different room configurations.
PPPR captures human-related information through kinematic constraints (Sec.~\ref{sec:optimization}, biomechanical consistency independent of room layouts) and electromagnetic reconstruction (Eq.~\ref{eq:em_iou}, signal physics independent of scene-specific multipath). 

PPPR enables convenient deployment across diverse indoor environments without site-specific recalibration, eliminating costly per-room data collection and fine-tuning required by conventional approaches.

\subsubsection{Cross-Dataset Evaluation}

We also evaluate PPPR on the challenging cross-dataset HPE scenarios, where performance may be jointly affected by both environment and radar noise. Compared with environmental noise, radar noise is another important factor yet underexplored. To evaluate PPPR's potential against both environment and radar noise, we train the RETR \cite{yataka2024retr} HPE model on the MMVR dataset \cite{rahman2024mmvr}, and test it on the HuPR \cite{lee2023hupr} and XRF55 \cite{wang2024xrf55} datasets. For PPPR, we perform radar-specific calibration: during initialization, we adapt the bin-to-physical coordinate mapping according to target radar specifications (range resolution, angular FOV); during radar simulation, we apply the target-specific FFT configuration described in Sec.~\ref{sec:radar_sim} to reconstruct Heatmaps matching the observed format. This calibration step requires no learned parameter updates—only physical parameter adjustments.

\begin{table}[h]
\centering
\setlength{\tabcolsep}{6pt}
\begin{adjustbox}{width=0.6\textwidth,center}
\begin{tabular}{llcccc}
\hline
\textbf{Training Dataset} & \textbf{Test Dataset} & \textbf{Input Type} & \textbf{MAJPE} & \textbf{PA-MAJPE} \\ \hline
\multirow{2}{*}{MMVR} & \multirow{2}{*}{MMVR} & Heatmap & 88.7 & 81.2 \\
& & \textbf{PPPR} & \textbf{69.3} & \textbf{63.5} \\ \hline
\multirow{2}{*}{MMVR} & \multirow{2}{*}{XRF55} & Heatmap & 218.7 & 201.3 \\
& & \textbf{PPPR} & \textbf{85.9} & \textbf{78.4} \\ \hline
\multirow{2}{*}{MMVR} & \multirow{2}{*}{HuPR} & Heatmap & 203.4 & 187.9 \\
& & \textbf{PPPR} & \textbf{82.6} & \textbf{75.1} \\ \hline
\end{tabular}
\end{adjustbox}
\caption{Cross-dataset evaluation of PPPR with the HPE model (RETR) trained on MMVR and tested on XRF55/HuPR without retraining. The first row shows within-dataset reference performance.}
\label{cross_dataset}
\end{table}

Table \ref{cross_dataset} reveals a critical limitation of Heatmap-based inputs. When models trained on MMVR are tested on XRF55 or HuPR without retraining, the Heatmap-based HPE model exhibits severe performance degradation: 218.7mm MAJPE on XRF55 (146.6\% increase over within-dataset 88.7mm) and 203.4mm MAJPE on HuPR (129.3\% increase over within-dataset 88.7mm). This degradation stems from the combined effect of environment and hardware noise: MMVR (256×128, 3D Unified FFT) differs from XRF55 (256×128, Doppler+Angle FFT) and HuPR (64×64×8, 3D Unified FFT) in both tensor dimensions and FFT configurations. Heatmap representations encode these radar-specific characteristics directly, causing models trained on one signal distribution to fail when encountering incompatible formats. While tensor reshaping enables numerical compatibility, underlying feature distributions remain mismatched.

In contrast, the PPPR-based HPE model achieves 85.9mm MAJPE on XRF55 (24.0\% increase over within-dataset 69.3mm) and 82.6mm MAJPE on HuPR (19.2\% increase over within-dataset 69.3mm), reducing cross-dataset error by 132.8mm (61\% relative reduction) on XRF55 and by 120.8mm (59\% relative reduction) on HuPR compared to Heatmap input. This improvement is enabled by PPPR's parametric representation that decouples pose estimation from radar-specific signal formats. The radar simulation mechanism (Sec. 4.2) adapts to different FFT configurations through electromagnetic operators ($M_{\mathrm{range}}, M_{\mathrm{Dopp}}, M_{\mathrm{angle}}$) calibrated according to target radar specifications, enabling unified parameter optimization across different radar platforms without retraining neural network weights. By modeling noise characteristics across different datasets parametrically, PPPR achieves hardware adaptation through calibration alone rather than data-driven retraining.



\rev{These results demonstrate PPPR's adaptability across different radar hardware and environments, suggesting that a model trained on one radar can be deployed to others via calibration rather than data-driven retraining.}

\subsection{Ablation Study}

In this subsection, we conduct ablation studies to examine several key components in optimizing PPPR, namely, initialization components (position and velocity, denoted as $P$ and $V$ in Fig.~\ref{main_pipeline_single}), radar simulation operators ($M_{\mathrm{range}}, M_{\mathrm{Dopp}}, M_{\mathrm{angle}}, M_{\mathrm{atten}}$ in Sec.~\ref{sec:radar_sim}), reconstruction threshold $\tau_{\mathrm{pct}}$ (Sec.~\ref{sec:optimization}), and constraint weights ($w_{\mathrm{EM}}$ and $w_{\mathrm{kine}}$ in Eq.~\ref{eq:total_loss}). We perform ablations on single-person frames, isolating PPPR's core mechanisms.

\subsubsection{Position and Velocity Initialization}

We ablate the initialization effect of position module $P$ and velocity module $V$ by replacing their outputs with zeros (referred to as w/o Position ($P$) and w/o Velocity ($V$) in Table~\ref{module_ablation}), evaluating their contributions across three datasets and two HPE models.

\begin{table}[h]
\centering
\setlength{\tabcolsep}{4pt}
\begin{adjustbox}{width=0.5\textwidth,center}
\begin{tabular}{llccc}
\hline
\textbf{HPE Model} & \textbf{Configuration} & \textbf{MMVR} & \textbf{HuPR} & \textbf{XRF55} \\ \hline
\multirow{4}{*}{RETR \cite{yataka2024retr}} & Full (P+V) & \textbf{69.30} & \textbf{75.90} & \textbf{77.51} \\
& w/o Position ($P$) & 89.74 & 84.05 & 89.43 \\
& w/o Velocity ($V$) & 71.92 & 78.11 & 80.45 \\ \hline
\multirow{4}{*}{mmDiff \cite{fan2024diffusion}} & Full (P+V) & \textbf{67.90} & \textbf{73.98} & \textbf{70.47} \\
& w/o Position ($P$) & 85.03 & 81.39 & 87.71 \\
& w/o Velocity ($V$) & 68.71 & 74.60 & 76.42 \\ \hline
\end{tabular}
\end{adjustbox}
\caption{Ablation on initialization modules (MAJPE$\downarrow$).}
\label{module_ablation}
\end{table}

Table~\ref{module_ablation} presents ablation results across two HPE models and three datasets. Removing position initialization increases MAJPE by 8-20mm (10-29\% relative increase) across all configurations, while removing velocity initialization increases MAJPE by 2-6mm (3-8\% relative increase). These results suggest that spatial initialization anchors Gaussian primitives and constrains the optimization search space, while velocity initialization provides secondary refinement through Doppler-consistent motion priors.

\subsubsection{Radar Pipeline Simulation Modules}

We ablate the four physics-based electromagnetic operators (Sec.~4.2.1) in the differentiable radar simulation pipeline using RETR \cite{yataka2024retr} on MMVR, systematically removing each component to quantify its contribution.

 \begin{table}[h]
\centering
\setlength{\tabcolsep}{5pt}
\begin{adjustbox}{width=0.5\textwidth,center}
\begin{tabular}{lcc}
\hline
\textbf{Configuration} & \textbf{MAJPE} & \textbf{PA-MAJPE} \\ \hline
Full Pipeline & \textbf{64.71} & \textbf{58.91} \\ \hline
w/o Range ($M_{\mathrm{range}}$) & 90.30 & 84.17 \\
w/o Doppler ($M_{\mathrm{Dopp}}$) & 89.00 & 82.66 \\
w/o Array Phase ($M_{\mathrm{angle}}$) & 70.92 & 65.05 \\
w/o Path Attenuation ($M_{\mathrm{atten}}$) & 66.84 & 60.27 \\ \hline
\end{tabular}
\end{adjustbox}
\caption{Ablation on radar simulation operators. }
\label{re-render_ablation}
\end{table}


Table~\ref{re-render_ablation} presents ablation results on radar simulation operators. Removing range modulation $M_{\mathrm{range}}$ increases MAJPE by 25.59mm (from 64.71mm to 90.30mm), while removing Doppler modulation $M_{\mathrm{Dopp}}$ increases MAJPE by 24.29mm (from 64.71mm to 89.00mm). Removing array phase modulation $M_{\mathrm{angle}}$ increases MAJPE by 6.21mm (from 64.71mm to 70.92mm), while removing path attenuation $M_{\mathrm{atten}}$ shows minimal impact, increasing MAJPE by 2.13mm (from 64.71mm to 66.84mm).

Experimental results suggest that range and Doppler modulations play major roles in radar signal encoding, as removing either component substantially increases error. One possible reason is that $M_{\mathrm{range}}$ establishes spatial correspondence between joint positions and range bins, while $M_{\mathrm{Dopp}}$ captures motion dynamics through frequency shifts. On the other hand, path attenuation $M_{\mathrm{atten}}$ has the least impact, which may be due to short detection distances in indoor scenarios where attenuation effects are minimal. Array phase modulation $M_{\mathrm{angle}}$ contributes secondary spatial discrimination, suggesting that angular resolution refinement provides incremental improvements beyond core range-Doppler encoding.

\subsubsection{Reconstruction Threshold $\tau_{\mathrm{pct}}$}

We investigate reconstruction constraint threshold $\tau_{\mathrm{pct}}$ (Sec.~4.3) by varying top-percentile intensity masking from 5\% to 100\% using RETR \cite{yataka2024retr} on MMVR's single-person scenarios.

\begin{table}[h]
\centering
\setlength{\tabcolsep}{1pt}
\begin{adjustbox}{width=0.6\textwidth,center}
\begin{tabular}{lcccccc}
\hline
\multirow{2}{*}{\textbf{Threshold $\tau_{\mathrm{pct}}$}} & \multicolumn{2}{c}{\textbf{HuPR}} & \multicolumn{2}{c}{\textbf{MMVR}} & \multicolumn{2}{c}{\textbf{XRF55}} \\ \cline{2-7}
& \textbf{MAJPE} & \textbf{PA-MAJPE} & \textbf{MAJPE} & \textbf{PA-MAJPE} & \textbf{MAJPE} & \textbf{PA-MAJPE} \\ \hline
\textbf{10\% (Optimal)} & \textbf{73.74} & \textbf{67.70} & \textbf{71.08} & \textbf{65.17} & \textbf{72.00} & \textbf{68.71} \\ \hline
5\% & 75.01 & 71.43 & 76.82 & 70.45 & 75.96 & 72.65 \\
30\% & 87.13 & 81.61 & 80.08 & 74.86 & 77.14 & 73.40 \\
70\% & 88.03 & 81.95 & 80.09 & 74.90 & 77.11 & 73.67 \\
100\% (No masking) & 88.17 & 82.05 & 80.51 & 75.47 & 77.89 & 75.03 \\ \hline
\end{tabular}
\end{adjustbox}
\caption{Ablation on reconstruction threshold $\tau_{\mathrm{pct}}$. }
\label{threshold_ablation}
\end{table}

Table~\ref{threshold_ablation} presents reconstruction threshold ablation results. Optimal performance occurs at $\tau_{\mathrm{pct}}=10\%$ across all datasets, achieving 71.08-73.74mm MAJPE. Overly aggressive 5\% masking increases MAJPE by 2-6mm (1-8\% relative increase), suggesting that discarding too much intensity information removes critical motion signals. Excessive retention (30-100\%) increases MAJPE by 5-14mm (7-19\% relative increase), indicating that incorporating lower-intensity regions introduces environment-dependent clutter. The consistent optimal threshold across diverse datasets suggests that the top 10\% intensities correspond to strong radar returns from body segments, while lower intensities predominantly contain multipath reflections and noise. This indicates that electromagnetic constraint supervision naturally identifies informative signal components without radar-specific tuning.

\begin{figure}[h]
  \centering
  \includegraphics[width=\linewidth]{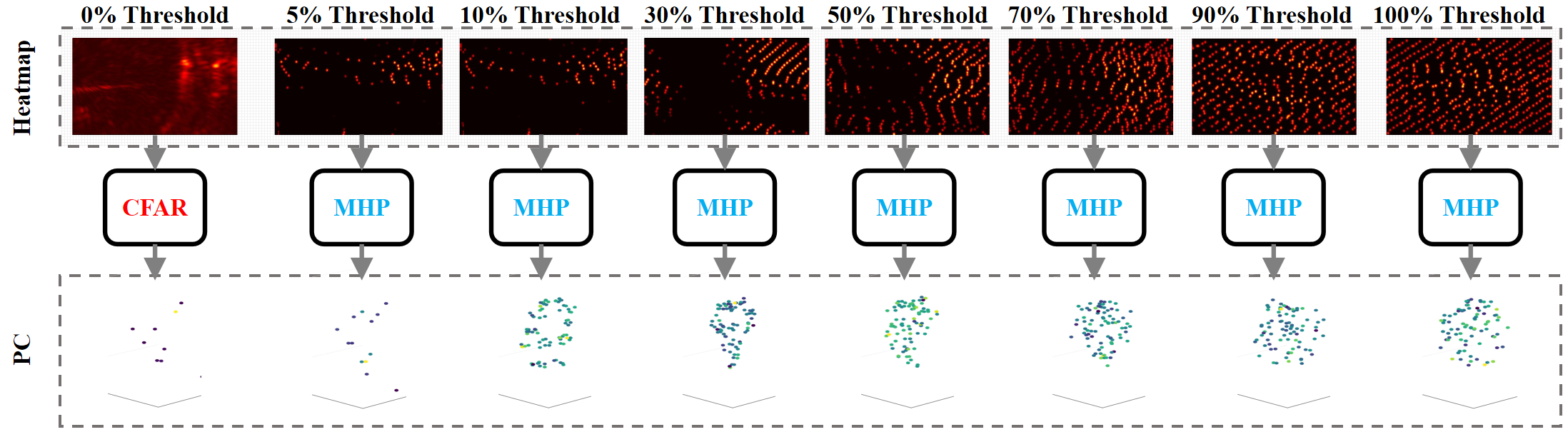}
  \caption{The impact of threshold on PPPR-enhanced Heatmap and PPPR-enhanced PC .} 
  \label{threshold} 
\end{figure}

\subsubsection{Electromagnetic and Kinematic Constraint Weights}

We examine electromagnetic and kinematic constraint weights $w_{\mathrm{EM}}$ and $w_{\mathrm{kine}}$ (Sec.~\ref{sec:optimization}, Eq.~\ref{eq:total_loss}) by varying their relative emphasis using RETR \cite{yataka2024retr} on single-person scenarios.

\begin{table}[h]
\centering
\setlength{\tabcolsep}{1pt}
\begin{adjustbox}{width=0.65\textwidth,center}
\begin{tabular}{lcccccc}
\hline
\multirow{2}{*}{\textbf{$w_{\mathrm{EM}}$ / $w_{\mathrm{kine}}$}} & \multicolumn{2}{c}{\textbf{HuPR}} & \multicolumn{2}{c}{\textbf{MMVR}} & \multicolumn{2}{c}{\textbf{XRF55}} \\ \cline{2-7}
& \textbf{MAJPE} & \textbf{PA-MAJPE} & \textbf{MAJPE} & \textbf{PA-MAJPE} & \textbf{MAJPE} & \textbf{PA-MAJPE} \\ \hline
\textbf{0.3 / 0.7} & \textbf{73.74} & \textbf{67.70} & 71.08 & 65.17 & \textbf{72.00} & \textbf{68.71} \\
\textbf{0.5 / 0.5} & 74.84 & 70.03 & \textbf{64.71} & \textbf{58.91} & 74.64 & 70.50 \\ \hline
0.7 / 0.3 & 75.31 & 72.47 & 75.08 & 72.33 & 77.09 & 74.63 \\ \hline
0.0 / 1.0 (Pure Kine) & 90.30 & 84.17 & 80.51 & 75.47 & 78.09 & 75.72 \\
1.0 / 0.0 (Pure Electro) & 80.88 & 75.93 & 78.73 & 74.11 & 78.09 & 75.72 \\ \hline
\end{tabular}
\end{adjustbox}
\caption{Ablation on electromagnetic and kinematic constraint weights.}
\label{constraint_ablation}
\end{table}

Table~\ref{constraint_ablation} presents ablation results on constraint weights. We examine five configurations: electromagnetic emphasis (0.3 / 0.7), balanced weighting (0.5 / 0.5), kinematic emphasis (0.7 / 0.3), pure kinematic optimization (0.0 / 1.0), and pure electromagnetic optimization (1.0 / 0.0).

Pure kinematic optimization increases MAJPE by 6-22mm across all datasets compared to optimal configurations, suggesting that biomechanical constraints alone without electromagnetic grounding cannot effectively constrain pose estimation. Pure electromagnetic optimization increases MAJPE by 7-11mm, indicating that signal consistency without biomechanical regularization yields implausible configurations. 
Optimal weight combinations exhibit dataset-dependent variation: MMVR achieves best performance with balanced weighting (0.5 / 0.5, 64.71mm MAJPE), while HuPR and XRF55 favor electromagnetic emphasis (0.3 / 0.7, achieving 73.74mm and 72.00mm MAJPE, respectively). These results suggest that MMVR's diverse multi-scene captures may benefit from stronger kinematic regularization to suppress environment-specific artifacts, while HuPR and XRF55's controlled environments enable reconstruction-driven optimization. This demonstrates that PPPR's dual-constraint architecture synergistically combines physics-grounded signal fidelity with biomechanical plausibility, with relative emphasis adaptable to dataset characteristics.

\subsection{Multi-Person HPE Scenarios}

\subsubsection{Count-Matched Evaluation Protocol}
We perform multi-person data quality control through a Count-Matched Evaluation Protocol (detailed in the Appendix~\ref{appendix:count_matched}).
Existing mmWave datasets (MMVR \cite{rahman2024mmvr}, HuPR \cite{lee2023hupr}, XRF55 \cite{wang2024xrf55}) exhibit certain levels of annotation errors for multi-person scenarios due to vision-based labeling systems. After performing this protocol, we can use the high-quality multi-person subset to evaluate PPPR-based HPE systems, which retains 91-95\% of original frames while ensuring annotation reliability.

\subsubsection{Person Counting}

\begin{table}[h]
\centering
\setlength{\tabcolsep}{4pt}
\begin{adjustbox}{width=0.9\textwidth,center}
\begin{tabular}{lccccc}
\hline
\textbf{Method} & \textbf{1-Person (91\%)} & \textbf{2-Person (8\%)} & \textbf{3-Person (1\%)} & \textbf{Weighted Acc.} & \textbf{FPS} \\ \hline

\rowcolor[HTML]{E6F3FF}
\textbf{ETCM-CFAR + DBSCAN + MLP} & \textbf{99.5\%} & \textbf{99.0\%} & \textbf{96.5\%} & \textbf{99.4\%} & \textbf{850} \\
Pure MLP & 95.0\% & 91.0\% & 83.0\% & 94.6\% & 1200 \\
CNN & 98.0\% & 93.5\% & 88.0\% & 97.5\% & 400 \\
ViT & 96.5\% & 94.0\% & 85.5\% & 96.2\% & 120 \\ \hline
\end{tabular}
\end{adjustbox}
\caption{Person counting accuracy on cleaned MMVR annotations (parentheses indicate frame distribution). Weighted accuracy computed as $\sum_{i=1}^{3} p_i \cdot \text{Acc}_i$ where $p_i$ denotes scenario prevalence.}
\label{tab:count_accuracy}
\end{table}

To extend PPPR to multi-person scenarios, we integrate a person counting module before per-person's PPPR parameters optimization. In the person counting module, we employ classical ETCM-CFAR \cite{rohling1983radar} thresholding followed by DBSCAN clustering \cite{ester1996density}, with an MLP head for count classification.

Table~\ref{tab:count_accuracy} presents person counting results on cleaned MMVR annotations. Parentheses indicate frame distribution: 1-person (91\%), 2-person (8\%), and 3-person (1\%) scenarios. The ETCM-CFAR + DBSCAN + MLP approach achieves 99.4\% weighted accuracy at 850 FPS, demonstrating superiority over pure MLP (94.6\%, 1200 FPS), CNN (97.5\%, 400 FPS), and ViT (96.2\%, 120 FPS) in both accuracy and efficiency. We use this classical approach as our person counting module, ensuring counting overhead remains negligible relative to pose optimization.

\begin{table}[h]
\centering
\label{tab:multiperson_indataset}
\begin{adjustbox}{width=0.9\textwidth,center}
\setlength{\tabcolsep}{4pt}
\begin{tabular}{ccccccc}
\hline
& & & \multicolumn{2}{c}{\textbf{MMVR}} & \multicolumn{2}{c}{\textbf{XRF55}} \\
\multirow{-2}{*}{\textbf{HPE Model}} & \multirow{-2}{*}{\textbf{Input}} & \multirow{-2}{*}{\textbf{Size}} & \textbf{MAJPE$\downarrow$} & \textbf{PA-MAJPE$\downarrow$} & \textbf{MAJPE$\downarrow$} & \textbf{PA-MAJPE$\downarrow$} \\ \hline

& Heatmap  & 76.9M & 81.39 & 72.51 & 86.14 & 83.06 \\
& PPPR & 12.9M & 72.93 (-8.46) & 70.26 (-2.25) & 83.15 (-2.99) & 79.47 (-3.59) \\
\multirow{-3}{*}{RETR \cite{yataka2024retr}} & PPPR-Heatmap & 76.9M & 80.26 (-1.13) & 69.57 (-2.94) & 84.09 (-2.05) & 80.07 (-2.99) \\
\cdashline{2-7}

& Heatmap  & 524.9M & 83.54 & 78.61 & 86.20 & 82.05 \\
& PPPR & 370.6M & 75.36 (-8.18) & 69.23 (-9.38) & 77.92 (-8.28) & 74.03 (-8.02) \\
\multirow{-3}{*}{HuprModel \cite{lee2023hupr}} & PPPR-Heatmap & 524.9M & 80.64 (-2.90) & 76.31 (-2.30) & 82.06 (-4.14) & 78.81 (-3.24) \\
\cdashline{2-7}

& PC  & 362.8M & 79.75 & 77.47 & 83.57 & 80.01 \\
& PPPR & 91.2M & 72.96 (-6.79) & 69.37 (-8.10) & 75.46 (-8.11) & 73.77 (-6.24) \\
\multirow{-3}{*}{mmDiff \cite{fan2024diffusion}} & PPPR-PC & 87.2M & 76.59 (-3.16) & 72.70 (-4.77) & 76.52 (-7.05) & 73.98 (-6.03) \\
\cdashline{2-7}

PoseformerV2 \cite{zhao2023poseformerv2} & PPPR  & 112.0M & 71.14 & 69.59 & 76.06 & 73.81 \\ \hline
\end{tabular}
\end{adjustbox}
\caption{Multi-person performance on manually verified 2-3 person subset.}
\end{table}

\subsubsection{Multi-Person In-Dataset Performance}

To evaluate PPPR-based systems in multi-person HPE settings, we examine performance on manually verified 2-3 person subsets, excluding single-person cases to highlight the challenges of multi-person scenarios. 

Table 9 presents multi-person results on 38,745 verified 2-3 person frames. PPPR achieves 6-9mm MAJPE improvements compared to baseline inputs across three HPE models on both MMVR and XRF55 datasets. Inter-person collision constraints ($\mathcal{L}_{\mathrm{sep}}, \mathcal{L}_{\mathrm{coll}}$) prevent skeletal overlap during optimization. Parametric modeling enables per-person signal decomposition: each PPPR instance optimizes independently under shared electromagnetic constraints while collision penalties enforce spatial separation.

\subsubsection{Multi-Person Cross-Scene Evaluation}

We also evaluate PPPR-based multi-person HPE systems in cross-scene settings. Table~\ref{tab:cross_scene_multi} presents results where models trained on MMVR Scene 1 are evaluated on Scenes 2-5.

\begin{table}[h]
\centering
\setlength{\tabcolsep}{2pt}
\begin{adjustbox}{width=0.6\textwidth,center}
\begin{tabular}{lcccccc}
\hline
\multirow{2}{*}{\textbf{Test Scene}} & \multicolumn{2}{c}{\textbf{mmDiff}} & \multicolumn{2}{c}{\textbf{HuPRmodel}} & \multicolumn{2}{c}{\textbf{RETR}} \\ \cline{2-7}
& \textbf{PC\cite{fan2024diffusion}} & \textbf{PPPR} & \textbf{Heatmap\cite{lee2023hupr}} & \textbf{PPPR} & \textbf{Heatmap\cite{yataka2024retr}} & \textbf{PPPR} \\ \hline
Scene 2  & 118.5 & 95.2 & 125.3 & 98.7 & 112.4 & 91.6 \\
Scene 3  & 119.2 & 96.0 & 126.1 & 99.3 & 113.1 & 92.3 \\
Scene 4  & 118.9 & 95.6 & 124.8 & 98.5 & 112.8 & 91.9 \\
Scene 5  & 120.1 & 96.4 & 125.7 & 99.1 & 113.5 & 92.5 \\ \hline
\textbf{Mean} & \textbf{119.2} & \textbf{95.8} & \textbf{125.5} & \textbf{98.9} & \textbf{113.0} & \textbf{92.1} \\
 \hline
\end{tabular}
\end{adjustbox}
\caption{Multi-person cross-scene MAJPE (mm) under count-matched evaluation. Training on Scene 1, testing on 2 to 5.}
\label{tab:cross_scene_multi}
\end{table}

\begin{figure}[h]
  \centering
  \includegraphics[width=0.9\linewidth]{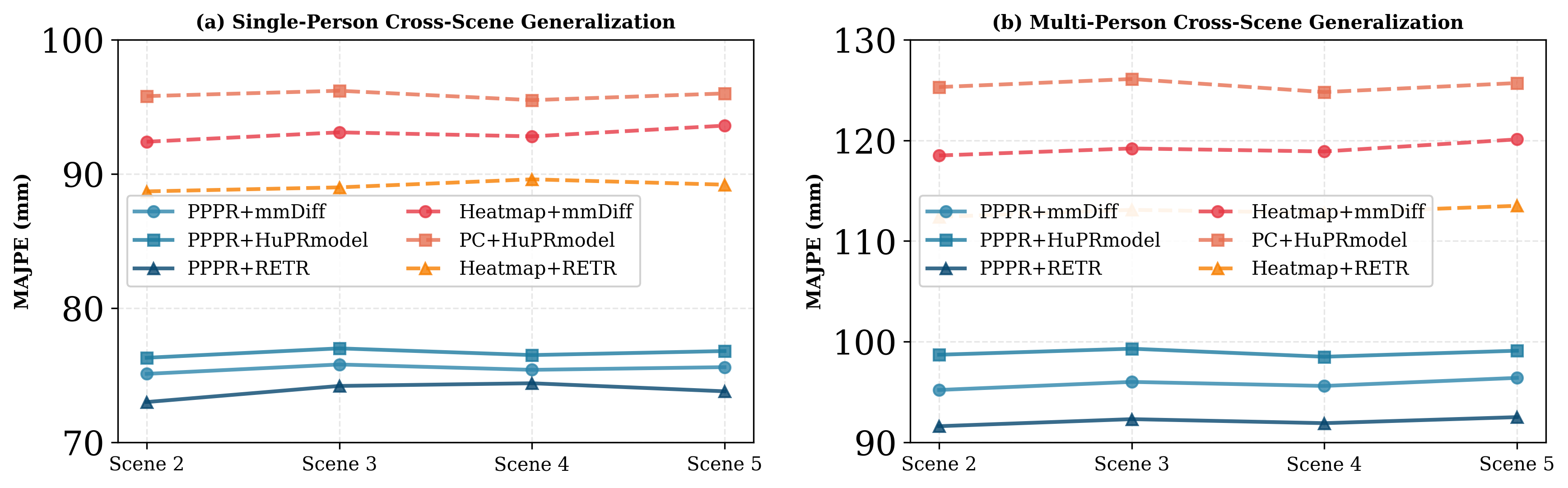}
  \caption{Cross-scene evaluation on single-person and multi-person settings.} 
  \label{cross_scene_generalization} 
\end{figure}

Table~\ref{tab:cross_scene_multi} presents cross-scene evaluation on MMVR multi-person scenarios. PPPR achieves consistent error reduction (19-24mm absolute, 16-20\% relative) compared to baseline inputs across all test scenes. Mean MAJPE improvements are 23.4mm (19.6\%) for mmDiff, 26.6mm (21.2\%) for HuPRmodel, and 20.9mm (18.5\%) for RETR. This demonstrates PPPR's capability to decouple human features from environment-induced noise in multi-person occlusion scenarios.

Figure~\ref{cross_scene_generalization} visualizes cross-scene performance trends intuitively. The blue-toned broken line represents the performance of the HPE system using PPPR as input, while the other colors represent the performance of the HPE system using Heatmap or PC as input. These performance differences confirm consistent improvements across scene variations.

\subsection{\rev{Complex-environment validation and failure analysis}}
\subsubsection{\rev{Complex-environment validation}}
\label{sec:mmvr_complex_validation}

\rev{To assess the behavior of the forward model and optimization in cluttered environments, we leverage MMVR's \cite{rahman2024mmvr} multi-room setting and evaluate PPPR under a reproducible complexity-stratified protocol. }

\rev{Each test frame is characterized by three label-agnostic statistics computed from the observed Heatmap $H_{\mathrm{ori}}$: (i) peak density $\downarrow$ $\rho_{\mathrm{peak}}$, defined as the number of local maxima above a high-intensity threshold (top-$\kappa$ percentile) normalized by Heatmap area; 
(ii) static dominance $\downarrow$ $s_{\mathrm{static}}$, defined as the temporal persistence of top-intensity bins over a short window of $T$ frames; and 
(iii) human-saliency ratio $\uparrow$ $r_{\mathrm{human}}=\sum(H_{\mathrm{ori}}\odot \mathcal{M}_{\tau})/\sum(H_{\mathrm{ori}})$, where $\mathcal{M}_{\tau}$ is the top-$\tau_{\mathrm{pct}}$ mask used in Eq.~\ref{eq:em_iou}. }

\begin{figure}[h]
\centering
\includegraphics[width=0.8\linewidth]{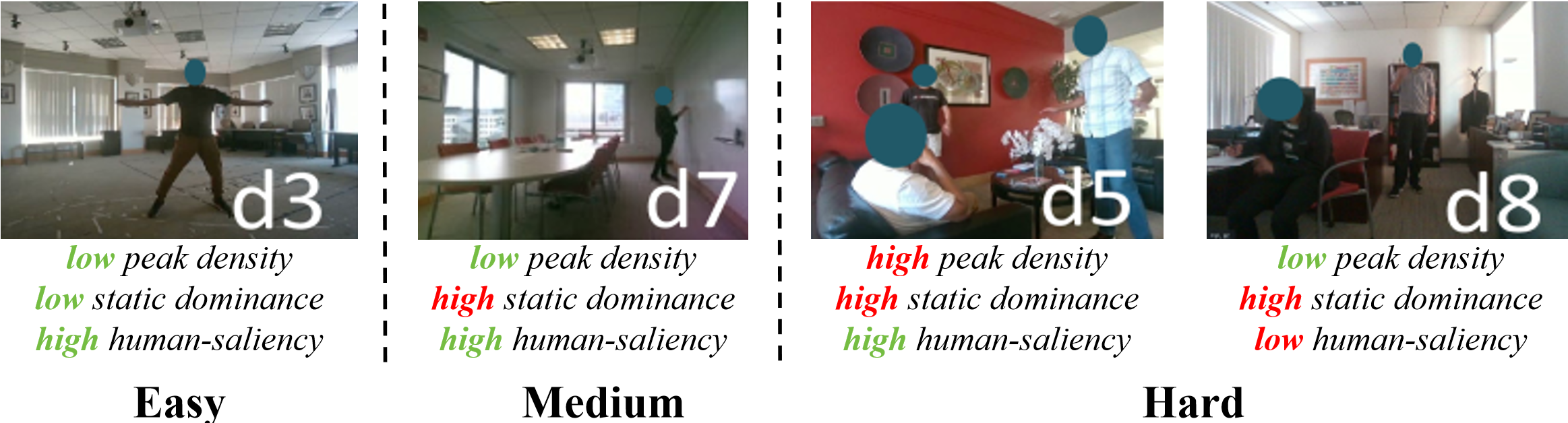}

\caption{\rev{Scenarios with varying environmental complexity, overlap of human body regions, and intensity of movement.}}

\label{fig:different_scene}
\end{figure}

\rev{We then split test frames in Scenes 2-5 into \textbf{easy/medium/hard} by quantiles as illustrated in Fig.~\ref{fig:different_scene}, making complex environments measurable and reproducible.
In addition to MAJPE/PA-MAJPE, we report \textbf{forward-consistency IoU}, computed post-optimization between the simulated Heatmap $H_{\mathrm{sim}}$ and observed $H_{\mathrm{ori}}$ using the same masking rule as Eq.~\ref{eq:em_iou}. This metric quantifies post-optimization electromagnetic consistency under clutter. We also report a convergence rate under a fixed iteration budget.}

\begin{table}[h]
\centering
\setlength{\tabcolsep}{12pt}
\begin{adjustbox}{width=0.85\linewidth,center}
\begin{tabular}{l|c|cc|cc|cc|c}
\hline
\multirow{2}{*}{\textbf{Subset}} 
& \multirow{2}{*}{\textbf{Frame}} 
& \multicolumn{2}{c|}{\textbf{mmDiff}} 
& \multicolumn{2}{c|}{\textbf{HuPRmodel}} 
& \multicolumn{2}{c|}{\textbf{RETR}}
& \multirow{2}{*}{\textbf{IoU/Conv$\uparrow$}} \\ \cline{3-8}
& 
& \textbf{PC$\downarrow$} & \textbf{PPPR$\downarrow$} 
& \textbf{Heatmap$\downarrow$} & \textbf{PPPR$\downarrow$} 
& \textbf{Heatmap$\downarrow$} & \textbf{PPPR$\downarrow$}
&  \\ \hline
Easy   & 71K  & 89.4 & 75.1 & 94.8 & 76.0 & 82.7 & 69.0 & 0.72 / 96.4\% \\
Medium & 68K & 93.1 & 75.8 & 94.9 & 76.3 & 88.0 & 72.2 & 0.71 / 96.0\% \\
Hard   & 61K & 97.2 & 75.6 & 98.5 & 77.8 & 97.8 & 81.5 & 0.69 / 94.8\% \\ \hline
\textbf{Mean} & - & \textbf{93.0} & \textbf{75.5} & \textbf{95.9} & \textbf{76.7} & \textbf{89.1} & \textbf{73.9} & \textbf{0.71 / 95.7\%} \\ \hline
\end{tabular}
\end{adjustbox}
\caption{\rev{MMVR complexity-stratified evaluation on Scenes 2--5. Columns Heatmap/PC/PPPR report \textbf{MAJPE (mm)} for each backbone. IoU/Conv reports post-optimization forward-consistency IoU (Eq.~\ref{eq:em_iou}) and convergence rate within the iteration budget; these quantities characterize PPPR optimization and are independent of the downstream backbone. Subsets are defined by quantiles of $(\rho_{\mathrm{peak}}, s_{\mathrm{static}}, r_{\mathrm{human}})$ with parameters $\kappa=5\%$, $T=5$, and $\tau_{\mathrm{pct}}=10\%$ (Sec.~\ref{sec:optimization}).}}
\label{tab:mmvr_complex_env}
\end{table}

\rev{Table~\ref{tab:mmvr_complex_env} shows that conventional inputs are sensitive to environmental complexity: as the subset shifts from easy to hard, MAJPE increases for PC/Heatmap baselines (e.g., mmDiff PC: 89.4$\rightarrow$97.2 mm; HuPRmodel Heatmap: 94.8$\rightarrow$98.5 mm; RETR Heatmap: 82.7$\rightarrow$97.8 mm). }

\rev{Replacing the input with PPPR reduces this sensitivity. For mmDiff and HuPRmodel, PPPR MAJPE varies only slightly across easy/medium/hard. For RETR, PPPR yields large gains on the easy/medium subsets (69.0/72.2 mm vs. 82.7/88.0 mm), while the hard subset remains more challenging (80.5 mm), indicating residual ambiguity under extreme clutter for certain backbones. }

\rev{Meanwhile, forward-consistency IoU and convergence remain more stable across subsets, suggesting that the electromagnetic consistency term remains informative and the optimization remains numerically stable as clutter increases.}

\rev{\textbf{Limitation.} While MMVR covers diverse indoor rooms with varying levels of clutter, it does not fully represent extremely strong-reflector regimes (e.g., pervasive metallic surfaces in industrial sites). Evaluating and extending PPPR under such conditions is left for future work.}

\begin{figure}[h]
\centering
\includegraphics[width=0.8\linewidth]{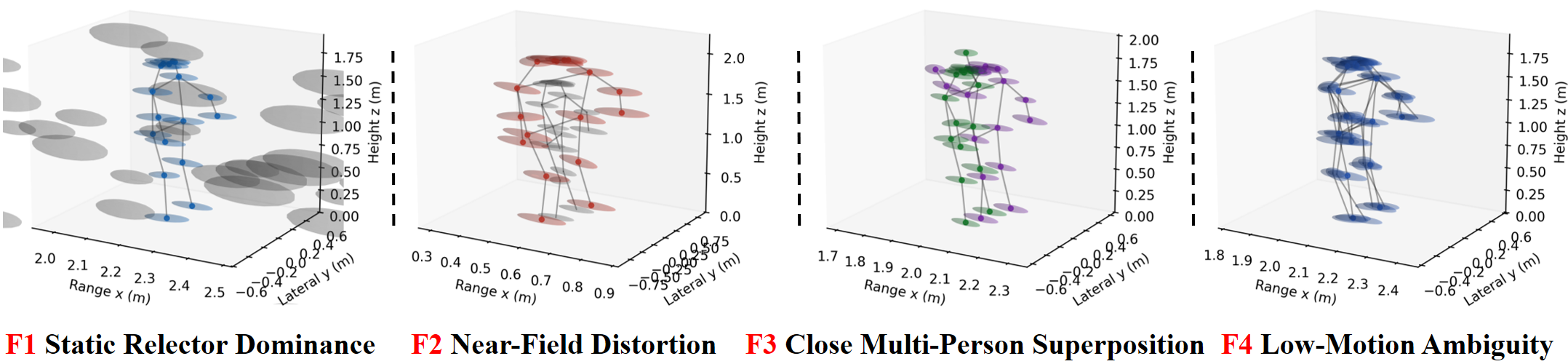}

\caption{\rev{Representative failure cases on the MMVR hard subset, illustrating four recurring modes: (F1) static reflector dominance, (F2) near-field distortion, (F3) close multi-person superposition, and (F4) low-motion ambiguity.}}

\label{fig:mmvr_failures}
\end{figure}

\subsubsection{\rev{Failure taxonomy.}} 
\rev{On the hard subset, we summarize four recurring failure modes: (F1) static reflector dominance, where high-energy masks are dominated by stationary structures; (F2) near-field distortion at very short ranges; (F3) close multi-person superposition; and (F4) low-motion ambiguity with weak Doppler cues. Fig.~\ref{fig:mmvr_failures} illustrates these regimes; in practice, they manifest as characteristic mismatches between $H_{\mathrm{sim}}$ and $H_{\mathrm{ori}}$ that can lead to large pose errors.}

\rev{We hypothesize that these failures are mainly driven by a few bottlenecks in MHP. (F1) can occur when the top-$\tau_{\mathrm{pct}}$ masking used by the electromagnetic IoU loss (Eq.~\ref{eq:em_iou}) is dominated by persistent static reflectors, weakening human identifiability; (F2) can be amplified by near-field phase/amplitude deviations that are not fully captured by the forward model (Sec.~\ref{sec:radar_sim}); (F3) reflects signal superposition and imperfect instance disentanglement when people are close; and (F4) arises when weak Doppler cues reduce the discriminative power of the electromagnetic objective and increase sensitivity to initialization. \rev{Mitigation directions are summarized in the Discussion (Sec.~5.6).}}

\subsubsection{\rev{Initialization robustness under clutter.}} 
\rev{Since MHP is original based on Heatmap-derived seeds, we evaluate whether PPPR optimization remains stable under biased initialization on the hard subset. We run controlled perturbation sweeps: (i) \textit{position perturbation} with Gaussian noise $\epsilon_p\!\sim\!\mathcal{N}(0,\sigma_p^2)$ and optional systematic bias $\Delta\mathbf{p}$, (ii) \textit{velocity corruption} with either $\mathbf{v}=\mathbf{0}$ or additive noise $\epsilon_v\!\sim\!\mathcal{N}(0,\sigma_v^2)$, and (iii) \textit{peak dropout} that randomly removes a fraction $r_{\mathrm{drop}}$ of detected peaks before clustering to mimic low-SNR missed detections. We report MAJPE/PA-MAJPE and convergence rate under a fixed iteration budget of $N_{\mathrm{iter}}=\textbf{80}$.}

\begin{table}[h]
\centering
\setlength{\tabcolsep}{5pt}
\begin{adjustbox}{width=0.7\linewidth,center}
\begin{tabular}{llcccc}
\hline
\textbf{Subset} & \textbf{Perturbation} & \textbf{Level} & \textbf{MAJPE$\downarrow$} & \textbf{PA-MAJPE$\downarrow$} & \textbf{Conv$\uparrow$.} \\
\hline
\multirow{3}{*}{Hard} & \multirow{3}{*}{$\mathbf{p}$ noise/bias} 
 & $\sigma_p=0.02$ m,\ $\Delta\mathbf{p}=0$ m      & \textbf{80.2} & \textbf{73.1} & \textbf{94.2\%} \\
 &  & $\sigma_p=0.05$ m,\ $\Delta\mathbf{p}=0.05$ m & \textbf{82.8} & \textbf{75.4} & \textbf{92.1\%} \\
 &  & $\sigma_p=0.08$ m,\ $\Delta\mathbf{p}=0.10$ m & \textbf{85.6} & \textbf{77.9} & \textbf{89.3\%} \\
\hline
\multirow{3}{*}{Hard} & \multirow{3}{*}{$\mathbf{v}$ corruption} 
 & $\mathbf{v}=\mathbf{0}$                          & \textbf{83.1} & \textbf{75.8} & \textbf{91.0\%} \\
 &  & $\sigma_v=0.15$ m/s                           & \textbf{82.2} & \textbf{75.0} & \textbf{92.4\%} \\
 &  & $\sigma_v=0.30$ m/s                           & \textbf{83.9} & \textbf{76.6} & \textbf{90.2\%} \\
\hline
\multirow{3}{*}{Hard} & \multirow{3}{*}{Peak dropout} 
 & $r_{\mathrm{drop}}=0.10$                         & \textbf{81.4} & \textbf{74.2} & \textbf{93.6\%} \\
 &  & $r_{\mathrm{drop}}=0.20$                      & \textbf{83.3} & \textbf{75.9} & \textbf{91.7\%} \\
 &  & $r_{\mathrm{drop}}=0.40$                      & \textbf{86.7} & \textbf{78.5} & \textbf{88.8\%} \\
\hline
Hard & Warm-start (online) & $\sigma_p=0.05$ m,\ $r_{\mathrm{drop}}=0.20$ & \textbf{79.6} & \textbf{72.6} & \textbf{95.3\%} \\
\hline
\end{tabular}
\end{adjustbox}
\caption{\rev{Initialization robustness on MMVR hard subset with RETR. ``Level'' specifies perturbation strength (e.g., $\sigma_p$, $\Delta\mathbf{p}$, $\sigma_v$, or $r_{\mathrm{drop}}$) and ``Conv.'' denotes convergence rate within the fixed iteration budget.}}
\label{tab:init_robustness}
\end{table}

\rev{The results in Table~\ref{tab:init_robustness} show that PPPR optimization remains stable on the MMVR hard subset under substantial initialization perturbations. As perturbation strength increases, MAJPE/PA-MAJPE increases and convergence decreases (e.g., convergence drops from 94.2\% to 89.3\% under larger $\sigma_p$ and $\Delta\mathbf{p}$, and from 93.6\% to 88.8\% when $r_{\mathrm{drop}}$ increases to 0.40), without abrupt failure across the tested range. This behavior is consistent with the optimization being guided by both kinematic regularization and electromagnetic consistency within a fixed iteration budget.}

\rev{We also observe that corrupting the velocity prior has a smaller impact than position bias or peak dropout: setting $\mathbf{v}=\mathbf{0}$ or adding moderate velocity noise keeps convergence at around 90–92\% with limited increase in error. This aligns with the role of velocity as a helpful but not exclusive cue, while geometric initialization and peak evidence more directly determine the initial correspondence between human structure and salient heatmap returns. Finally, the online warm-start setting further improves both accuracy and convergence (79.6/72.6 mm at 95.3\%), highlighting that temporal continuity can partially compensate for imperfect per-frame initialization and making the overall pipeline more reliable under realistic streaming deployment.}

\subsubsection{\rev{Convergence criterion.}} 
\rev{We define convergence to compute the convergence rate reported in Sec .~\ref {sec:mmvr_complex_validation}. A trial is counted as converged if (a) the total loss (Eq.~\ref{eq:total_loss}) decreases below $\epsilon_{\mathcal{L}}=\textbf{1.0}$ or its relative improvement over the last $\Delta t=\textbf{5}$ iterations is below $\epsilon_{\mathrm{rel}}=\textbf{0.5\%}$, and (b) the resulting skeleton is non-degenerate: all joint coordinates are finite (no NaN/Inf), and for all bones $(m,n)\!\in\!\mathcal{E}$, $\|\mathbf{p}_m-\mathbf{p}_n\|\in[0.5\,\ell_{mn},\,1.5\,\ell_{mn}]$. Runs that do not satisfy these conditions are treated as non-converged for the convergence statistics. We also report an online warm-start setting (using previous-frame PPPR parameters as initialization) to reflect streaming deployment.
}

\subsection{Discussion}

This section discusses practical deployment considerations, computational efficiency, and limitations of our work.

\subsubsection{Computational Efficiency}

Table~\ref{mlp} compares computational efficiency between PC-based and PPPR-based configurations using the mmDiff \cite{fan2024diffusion} HPE model on the MMVR dataset.

\begin{table}[h]
\centering
\setlength{\tabcolsep}{2pt}
\begin{adjustbox}{width=0.6\textwidth,center}
\begin{tabular}{lcccccc}
\hline
\textbf{Configuration} & \textbf{Params$\downarrow$} & \textbf{Latency$\downarrow$} & \textbf{FPS$\uparrow$} & \textbf{Memory$\downarrow$} & \textbf{MAJPE$\downarrow$} & \textbf{PA-MAJPE$\downarrow$} \\ \hline
mmDiff+PC & 362.8M & 24ms & 41.7 & 8.0GB & 68.62 & 62.11 \\ \hline
mmDiff+PPPR & 91.2M & 11ms & 90.9 & 3.6GB & \textbf{61.57} & \textbf{54.99} \\ \hline
MLP+PPPR & 3.9M & 4ms & 250.0 & 0.5GB & 72.50 & 69.11 \\ \hline

\end{tabular}
\end{adjustbox}
\caption{\rev{Computational efficiency comparison on MMVR single-person scenarios. Results are reported as model parameters, inference latency in milliseconds, frames per second, GPU memory in gigabytes, and MAJPE/PA-MAJPE in millimeters. The MLP architecture used by the ``MLP+PPPR'' row is reported in Appendix~\ref{appendix:mlp_pppr_arch}. }}

\label{mlp}
\end{table}

Table~\ref{mlp} presents computational efficiency analysis. Integrating PPPR into mmDiff reduces model parameters by 75\% (from 362.8M to 91.2M), inference latency by 54\% (from 24ms to 11ms), and GPU memory by 55\% (from 8.0GB to 3.6GB), while doubling FPS from 41.7 to 90.9 and improving MAJPE by 10.3\% (from 68.62mm to 61.57mm). A lightweight MLP+PPPR configuration achieves 250 FPS at 3.9M parameters with acceptable accuracy (72.50mm MAJPE, 5.7\% degradation compared to baseline). \rev{We do not report an ``MLP+PC'' row since PC is a variable-length point set; making it compatible with an MLP requires additional point-set preprocessing/encoding, including sampling and permutation-invariant pooling or a dedicated point encoder, which would introduce extra modules beyond a plain MLP.}

\rev{For deployment considerations, we further report an end-to-end latency breakdown of the full pipeline (initialization + radar simulation + iterative optimization + backbone inference) and an online warm-start setting in Appendix~\ref{appendix:end_to_end_runtime}.}

This efficiency gain arises from PPPR's compact parametric representation (2.8KB per person) that eliminates redundant noise channels present in dense Heatmaps (256KB) and sparsity-induced overhead in Point Clouds. Denoised inputs simplify downstream HPE model learning, enabling smaller models without compromising accuracy.

\subsubsection{PPPR's High Signal-to-Noise Ratio Property}
\rev{PPPR is designed to improve input signal-to-noise ratio (SNR) by compressing raw radar observations into a physics-informed parametric state, then refining it via electromagnetic and kinematic constraints (Sec.~4.1 and Sec.~\ref{sec:optimization}). Empirically, PPPR consistently improves accuracy and robustness relative to conventional Heatmap/PC inputs across within-dataset, cross-scene, and cross-dataset settings (Sec.~5.2--5.3). We therefore interpret PPPR as a practical intermediate representation that reduces environment-/hardware-induced nuisance factors while preserving human-relevant structure, which helps downstream HPE models generalize under clutter.}

\rev{Importantly, PPPR does not claim independence from Heatmap evidence: MHP explicitly constrains PPPR parameters by matching the observed Heatmap through electromagnetic-consistency objectives (Eq.~\ref{eq:em_iou} and Eq.~\ref{eq:total_loss}). Our contribution is to make this evidence use more structured and physically grounded, by fitting within an explicit human-centric parameter space and physics-informed constraints, rather than treating Heatmaps as the final learned input.}

\subsubsection{\rev{Limitations and future work}}
\rev{We summarize several directions that are motivated by real-world applications and by the failure regimes observed in complex environments (Sec.~\ref{sec:mmvr_complex_validation}):}
\begin{itemize}
    \item \rev{\textbf{Stronger real-world perturbations.} Current public benchmarks do not fully cover extreme strong-reflector scenes (e.g., large metallic surfaces) or external interference sources. A natural next step is to build/curate evaluation sets with such perturbations and report robustness under controlled stress tests.}
    \item \rev{\textbf{Denser human representations beyond joints.} PPPR currently parameterizes human pose at the joint level. Extending the same physics-informed parameterization idea to denser representations (e.g., mesh or surface primitives) may benefit broader human sensing tasks beyond HPE, but requires new supervision and careful modeling of scattering/occlusion.}
    \item \rev{\textbf{Deployment on edge devices.} While we report end-to-end latency breakdowns and online warm-start settings (Appendix~\ref{appendix:end_to_end_runtime}), practical deployment also depends on power/energy and platform constraints. Future work will benchmark PPPR+MHP on representative edge hardware and study accuracy--latency--energy trade-offs under reduced iteration budgets and quantized/low-precision inference.}
    \item \rev{\textbf{Generalization to other articulated targets.} Our focus is human pose, but the core idea---physics-informed parameterization with forward-consistency---may extend to other articulated targets (e.g., animals, robots). We leave such generalization studies to future work.}
\end{itemize}

\section{Conclusion}



\rev{This work targets a central challenge in mmWave-based human pose estimation: pervasive noise from environmental reflections and radar processing artifacts. We propose Person Parametric Physics-informed Representation (PPPR), a physics-informed parametric intermediate representation refined by electromagnetic and kinematic constraints. Across multiple backbones and evaluation settings, PPPR improves pose accuracy and robustness in cluttered scenes and across datasets, suggesting improved deployability under realistic indoor conditions. Future work will extend stress-test evaluation to stronger-reflector/interference-heavy environments and study deployment on resource-constrained platforms.}

\bibliographystyle{abbrv}
\bibliography{bibliography}

@STRING{jan = "Jan."}

@STRING{health = "ACM Transactions on Computing for Healthcare"}

@article{ionescu2013human3,
  title={Human3. 6m: Large scale datasets and predictive methods for 3d human sensing in natural environments},
  author={Ionescu, Catalin and Papava, Dragos and Olaru, Vlad and Sminchisescu, Cristian},
  journal={IEEE transactions on pattern analysis and machine intelligence},
  volume={36},
  number={7},
  pages={1325--1339},
  year={2013},
  publisher={IEEE}
}

@inproceedings{akhter2015pose,
  title={Pose-conditioned joint angle limits for 3D human pose reconstruction},
  author={Akhter, Ijaz and Black, Michael J},
  booktitle={Proceedings of the IEEE conference on computer vision and pattern recognition},
  pages={1446--1455},
  year={2015}
}

@inproceedings{jiang2022neuman,
  title={Neuman: Neural human radiance field from a single video},
  author={Jiang, Wei and Yi, Kwang Moo and Samei, Golnoosh and Tuzel, Oncel and Ranjan, Anurag},
  booktitle={European Conference on Computer Vision},
  pages={402--418},
  year={2022},
  organization={Springer}
}

@book{van2002optimum,
  title={Optimum array processing: Part IV of detection, estimation, and modulation theory},
  author={Van Trees, Harry L},
  year={2002},
  publisher={John Wiley \& Sons}
}

@book{richards2005fundamentals,
  title={Fundamentals of radar signal processing},
  author={Richards, Mark A and others},
  volume={1},
  year={2005},
  publisher={Mcgraw-hill New York}
}

@inproceedings{piersanti2012millimeter,
  title={Millimeter waves channel measurements and path loss models},
  author={Piersanti, Stefano and Annoni, Luca Alfredo and Cassioli, Dajana},
  booktitle={2012 IEEE International Conference on Communications (ICC)},
  pages={4552--4556},
  year={2012},
  organization={IEEE}
}

@article{sola2017quaternion,
  title={Quaternion kinematics for the error-state Kalman filter},
  author={Sola, Joan},
  journal={arXiv preprint arXiv:1711.02508},
  year={2017}
}

@article{alfarano2024estimating,
  title={Estimating optical flow: A comprehensive review of the state of the art},
  author={Alfarano, Andrea and Maiano, Luca and Papa, Lorenzo and Amerini, Irene},
  journal={Computer vision and image understanding},
  volume={249},
  pages={104160},
  year={2024},
  publisher={Elsevier}
}

@article{zhu2025probradarm3f,
  title={ProbRadarM3F: mmWave Radar-based Human Skeletal Pose Estimation with Probability Map Guided Multi-Format Feature Fusion},
  author={Zhu, Bing and He, Zixin and Xiong, Weiyi and Ding, Guanhua and Huang, Tao and Xiang, Wei},
  journal={IEEE Transactions on Aerospace and Electronic Systems},
  year={2025},
  publisher={IEEE}
}

@inproceedings{choi2025mvdoppler,
  title={MVDoppler-Pose: Multi-Modal Multi-View mmWave Sensing for Long-Distance Self-Occluded Human Walking Pose Estimation},
  author={Choi, Jaeho and Hor, Soheil and Yang, Shubo and Arbabian, Amin},
  booktitle={Proceedings of the Computer Vision and Pattern Recognition Conference},
  pages={27750--27759},
  year={2025}
}

@article{zhang2025breaking,
  title={Breaking the Resolution Barriers of mmWave Arrays via Null Steering for Sleep Monitoring in Multi-Person Scenarios},
  author={Zhang, Duo and Zhang, Xusheng and Yin, Zhehui and Zhou, Pengfei and Yang, Hongliu and Wang, Pei and Yao, Zhiyun and Wang, Junzhe and Zhang, Fusang and Zhang, Daqing},
  journal={Proceedings of the ACM on Interactive, Mobile, Wearable and Ubiquitous Technologies},
  volume={9},
  number={3},
  pages={1--29},
  year={2025},
  publisher={ACM New York, NY, USA}
}

@inproceedings{yang2025iradar,
  title={iRadar: Synthesizing Millimeter-Waves from Wearable Inertial Inputs for Human Gesture Sensing},
  author={Yang, Huanqi and Han, Mingda and Li, Xinyue and Duan, Di and Li, Tianxing and Xu, Weitao},
  booktitle={IEEE INFOCOM 2025-IEEE Conference on Computer Communications},
  pages={1--10},
  year={2025},
  organization={IEEE}
}

@article{feng20243d,
  title={3D Bounding Box Estimation Based on COTS mmWave Radar via Moving Scanning},
  author={Feng, Yiwen and Zhao, Jiayang and Wang, Chuyu and Xie, Lei and Lu, Sanglu},
  journal={Proceedings of the ACM on Interactive, Mobile, Wearable and Ubiquitous Technologies},
  volume={8},
  number={4},
  pages={1--27},
  year={2024},
  publisher={ACM New York, NY, USA}
}

@article{palipana2021pantomime,
  title={Pantomime: Mid-air gesture recognition with sparse millimeter-wave radar point clouds},
  author={Palipana, Sameera and Salami, Dariush and Leiva, Luis A and Sigg, Stephan},
  journal={Proceedings of the ACM on interactive, mobile, wearable and ubiquitous technologies},
  volume={5},
  number={1},
  pages={1--27},
  year={2021},
  publisher={ACM New York, NY, USA}
}

@article{mei2024mmspyvr,
  title={mmSpyVR: Exploiting mmWave radar for penetrating obstacles to uncover privacy vulnerability of virtual reality},
  author={Mei, Luoyu and Liu, Ruofeng and Yin, Zhimeng and Zhao, Qingchuan and Jiang, Wenchao and Wang, Shuai and Lu, Kangjie and He, Tian},
  journal={Proceedings of the ACM on Interactive, Mobile, Wearable and Ubiquitous Technologies},
  volume={8},
  number={4},
  pages={1--29},
  year={2024},
  publisher={ACM New York, NY, USA}
}

@article{qian20203d,
  title={3D point cloud generation with millimeter-wave radar},
  author={Qian, Kun and He, Zhaoyuan and Zhang, Xinyu},
  journal={Proceedings of the ACM on Interactive, Mobile, Wearable and Ubiquitous Technologies},
  volume={4},
  number={4},
  pages={1--23},
  year={2020},
  publisher={ACM New York, NY, USA}
}

@article{liu2024view,
  title={View-agnostic Human Exercise Cataloging with Single MmWave Radar},
  author={Liu, Alan and Lin, Yu-Tai and Sundaresan, Karthikeyan},
  journal={Proceedings of the ACM on Interactive, Mobile, Wearable and Ubiquitous Technologies},
  volume={8},
  number={3},
  pages={1--23},
  year={2024},
  publisher={ACM New York, NY, USA}
}

@article{cai2023millipcd,
  title={Millipcd: Beyond traditional vision indoor point cloud generation via handheld millimeter-wave devices},
  author={Cai, Pingping and Sur, Sanjib},
  journal={Proceedings of the ACM on Interactive, Mobile, Wearable and Ubiquitous Technologies},
  volume={6},
  number={4},
  pages={1--24},
  year={2023},
  publisher={ACM New York, NY, USA}
}

@inproceedings{chen2022mmbody,
  title={mmbody benchmark: 3d body reconstruction dataset and analysis for millimeter wave radar},
  author={Chen, Anjun and Wang, Xiangyu and Zhu, Shaohao and Li, Yanxu and Chen, Jiming and Ye, Qi},
  booktitle={Proceedings of the 30th ACM International Conference on Multimedia},
  pages={3501--3510},
  year={2022}
}

@inproceedings{ahuja2021vid2doppler,
  title={Vid2doppler: Synthesizing doppler radar data from videos for training privacy-preserving activity recognition},
  author={Ahuja, Karan and Jiang, Yue and Goel, Mayank and Harrison, Chris},
  booktitle={Proceedings of the 2021 CHI Conference on Human Factors in Computing Systems},
  pages={1--10},
  year={2021}
}

@inproceedings{chen2023rf,
  title={Rf genesis: Zero-shot generalization of mmwave sensing through simulation-based data synthesis and generative diffusion models},
  author={Chen, Xingyu and Zhang, Xinyu},
  booktitle={Proceedings of the 21st ACM Conference on Embedded Networked Sensor Systems},
  pages={28--42},
  year={2023}
}

@article{ren2024noncontact,
  title={Noncontact Multipoint Vital Sign Monitoring With mmWave MIMO Radar},
  author={Ren, Wei and Cao, Jiannong and Yi, Huansheng and Hou, Kaiyue and Hu, Miaoyang and Wang, Jianqi and Qi, Fugui},
  journal={IEEE Transactions on Microwave Theory and Techniques},
  year={2024},
  publisher={IEEE}
}

@article{han20234d,
  title={4d millimeter-wave radar in autonomous driving: A survey},
  author={Han, Zeyu and Wang, Jiahao and Xu, Zikun and Yang, Shuocheng and He, Lei and Xu, Shaobing and Wang, Jianqiang and Li, Keqiang},
  journal={arXiv preprint arXiv:2306.04242},
  year={2023}
}

@inproceedings{wang2019densefusion,
  title={Densefusion: 6d object pose estimation by iterative dense fusion},
  author={Wang, Chen and Xu, Danfei and Zhu, Yuke and Mart{\'\i}n-Mart{\'\i}n, Roberto and Lu, Cewu and Fei-Fei, Li and Savarese, Silvio},
  booktitle={Proceedings of the IEEE/CVF conference on computer vision and pattern recognition},
  pages={3343--3352},
  year={2019}
}

@article{xiang2017posecnn,
  title={Posecnn: A convolutional neural network for 6d object pose estimation in cluttered scenes},
  author={Xiang, Yu and Schmidt, Tanner and Narayanan, Venkatraman and Fox, Dieter},
  journal={arXiv preprint arXiv:1711.00199},
  year={2017}
}

@inproceedings{zhao2018through,
  title={Through-wall human pose estimation using radio signals},
  author={Zhao, Mingmin and Li, Tianhong and Abu Alsheikh, Mohammad and Tian, Yonglong and Zhao, Hang and Torralba, Antonio and Katabi, Dina},
  booktitle={Proceedings of the IEEE conference on computer vision and pattern recognition},
  pages={7356--7365},
  year={2018}
}

@inproceedings{cheng2021boundary,
  title={Boundary IoU: Improving object-centric image segmentation evaluation},
  author={Cheng, Bowen and Girshick, Ross and Doll{\'a}r, Piotr and Berg, Alexander C and Kirillov, Alexander},
  booktitle={Proceedings of the IEEE/CVF conference on computer vision and pattern recognition},
  pages={15334--15342},
  year={2021}
}

@article{toh2023usability,
  title={Usability of a wearable device for home-based upper limb telerehabilitation in persons with stroke: a mixed-methods study},
  author={Toh, Sharon Fong Mei and Gonzalez, Pablo Cruz and Fong, Kenneth NK},
  journal={Digital Health},
  volume={9},
  pages={20552076231153737},
  year={2023},
  publisher={SAGE Publications Sage UK: London, England}
}

@article{salehzadeh2024wearable,
  title={Wearable activity trackers: A survey on utility, privacy, and security},
  author={Salehzadeh Niksirat, Kavous and Velykoivanenko, Lev and Zufferey, No{\'e} and Cherubini, Mauro and Huguenin, K{\'e}vin and Humbert, Mathias},
  journal={ACM Computing Surveys},
  volume={56},
  number={7},
  pages={1--40},
  year={2024},
  publisher={ACM New York, NY}
}

@article{alshehri2022exploring,
  title={Exploring the privacy concerns of bystanders in smart homes from the perspectives of both owners and bystanders},
  author={Alshehri, Ahmed and Spielman, Joseph and Prasad, Amiya and Yue, Chuan},
  journal={Proceedings on Privacy Enhancing Technologies},
  year={2022}
}

@article{guhr2020privacy,
  title={Privacy concerns in the smart home context},
  author={Guhr, Nadine and Werth, Oliver and Blacha, Philip Peter Hermann and Breitner, Michael H},
  journal={SN Applied Sciences},
  volume={2},
  number={2},
  pages={247},
  year={2020},
  publisher={Springer}
}

@article{liu2024pmtrack,
  title={Pmtrack: Enabling personalized mmwave-based human tracking},
  author={Liu, Hankai and Liu, Xiulong and Xie, Xin and Tong, Xinyu and Li, Keqiu},
  journal={Proceedings of the ACM on Interactive, Mobile, Wearable and Ubiquitous Technologies},
  volume={7},
  number={4},
  pages={1--30},
  year={2024},
  publisher={ACM New York, NY, USA}
}

@article{engel2025advanced,
  title={Advanced Millimeter Wave Radar-Based Human Pose Estimation Enabled by a Deep Learning Neural Network Trained With Optical Motion Capture Ground Truth Data},
  author={Engel, Lukas and Mueller, Jonas and Rendon, Eduardo Javier Feria and Dorschky, Eva and Krauss, Daniel and Ullmann, Ingrid and Eskofier, Bjoern M and Vossiek, Martin},
  journal={IEEE Journal of Microwaves},
  year={2025},
  publisher={IEEE}
}

@inproceedings{weng2022humannerf,
  title={Humannerf: Free-viewpoint rendering of moving people from monocular video},
  author={Weng, Chung-Yi and Curless, Brian and Srinivasan, Pratul P and Barron, Jonathan T and Kemelmacher-Shlizerman, Ira},
  booktitle={Proceedings of the IEEE/CVF conference on computer vision and pattern Recognition},
  pages={16210--16220},
  year={2022}
}

@article{yang2025spatial,
  title={From Spatial Domain to Temporal Domain: Unleashing the Capability of CFAR for mmWave Point Cloud Generation},
  author={Yang, Hongliu and Zhang, Duo and Zhang, Xusheng and Xiong, Jie and Fan, Zizhou and Ning, Wanru and Chen, Weiyan and Zhang, Fusang and Han, Zijun and Zhang, Daqing},
  journal={Proceedings of the ACM on Interactive, Mobile, Wearable and Ubiquitous Technologies},
  volume={9},
  number={2},
  pages={1--29},
  year={2025},
  publisher={ACM New York, NY, USA}
}

@inproceedings{liu2025umotion,
  title={UMotion: Uncertainty-driven Human Motion Estimation from Inertial and Ultra-wideband Units},
  author={Liu, Huakun and Ota, Hiroki and Wei, Xin and Hirao, Yutaro and Perusquia-Hernandez, Monica and Uchiyama, Hideaki and Kiyokawa, Kiyoshi},
  booktitle={Proceedings of the Computer Vision and Pattern Recognition Conference},
  pages={7085--7094},
  year={2025}
}

@article{sengupta2020mm,
  title={mm-Pose: Real-time human skeletal posture estimation using mmWave radars and CNNs},
  author={Sengupta, Arindam and Jin, Feng and Zhang, Renyuan and Cao, Siyang},
  journal={IEEE sensors journal},
  volume={20},
  number={17},
  pages={10032--10044},
  year={2020},
  publisher={IEEE}
}

@article{hsieh2024multiperson,
  title={Multiperson localization and vital signs estimation using mmwave mimo radar},
  author={Hsieh, Chieh-Hsun and Tseng, Po-Hsuan},
  journal={IEEE Transactions on Microwave Theory and Techniques},
  year={2024},
  publisher={IEEE}
}

@inproceedings{wu2020mmtrack,
  title={mmTrack: Passive multi-person localization using commodity millimeter wave radio},
  author={Wu, Chenshu and Zhang, Feng and Wang, Beibei and Liu, KJ Ray},
  booktitle={IEEE INFOCOM 2020-IEEE Conference on Computer Communications},
  pages={2400--2409},
  year={2020},
  organization={IEEE}
}

@article{chen2006micro,
  title={Micro-Doppler effect in radar: phenomenon, model, and simulation study},
  author={Chen, Victor C and Li, Fayin and Ho, S-S and Wechsler, Harry},
  journal={IEEE Transactions on Aerospace and electronic systems},
  volume={42},
  number={1},
  pages={2--21},
  year={2006},
  publisher={IEEE}
}

@String{Computing = "Computing" }

@String{Computer = "{IEEE} Computer" }

@String{Springer = "Springer-Verlag" }

@BOOK{test,
   author = "Donald E. Knuth",
   title = "Seminumerical Algorithms",
   volume = 2,
   series = "The Art of Computer Programming",
   publisher = "Addison-Wesley",
   address = "Reading, MA",
   edition = "2nd",
   month = "10~" # jan,
   year = "1981",
}

@ArtifactSoftware{R,
    title = {R: A Language and Environment for Statistical Computing},
    author = {{R Core Team}},
    organization = {R Foundation for Statistical Computing},
    address = {Vienna, Austria},
    year = {2019},
    url = {https://www.R-project.org/},
}

@inproceedings{zhao2018rf,
  title={RF-based 3D skeletons},
  author={Zhao, Mingmin and Tian, Yonglong and Zhao, Hang and Alsheikh, Mohammad Abu and Li, Tianhong and Hristov, Rumen and Kabelac, Zachary and Katabi, Dina and Torralba, Antonio},
  booktitle={Proceedings of the 2018 Conference of the ACM Special Interest Group on Data Communication},
  pages={267--281},
  year={2018}
}

@article{sengupta2022mmpose,
  title={mmpose-nlp: A natural language processing approach to precise skeletal pose estimation using mmwave radars},
  author={Sengupta, Arindam and Cao, Siyang},
  journal={IEEE Transactions on Neural Networks and Learning Systems},
  volume={34},
  number={11},
  pages={8418--8429},
  year={2022},
  publisher={IEEE}
}

@article{chang2020spatial,
  title={Spatial attention fusion for obstacle detection using mmwave radar and vision sensor},
  author={Chang, Shuo and Zhang, Yifan and Zhang, Fan and Zhao, Xiaotong and Huang, Sai and Feng, Zhiyong and Wei, Zhiqing},
  journal={Sensors},
  volume={20},
  number={4},
  pages={956},
  year={2020},
  publisher={MDPI}
}

@article{cao2022joint,
  title={A joint global--local network for human pose estimation with millimeter wave radar},
  author={Cao, Zhongping and Ding, Wen and Chen, Rihui and Zhang, Jianxiong and Guo, Xuemei and Wang, Guoli},
  journal={IEEE Internet of Things Journal},
  volume={10},
  number={1},
  pages={434--446},
  year={2022},
  publisher={IEEE}
}

@article{an2022mri,
  title={mri: Multi-modal 3d human pose estimation dataset using mmwave, rgb-d, and inertial sensors},
  author={An, Sizhe and Li, Yin and Ogras, Umit},
  journal={Advances in neural information processing systems},
  volume={35},
  pages={27414--27426},
  year={2022}
}

@article{iovescu2020fundamentals,
  title={The fundamentals of millimeter wave radar sensors},
  author={Iovescu, Cesar and Rao, Sandeep},
  journal={Texas Instruments},
  pages={1--7},
  year={2020}
}

@article{wang2024xrf55,
  title={Xrf55: A radio frequency dataset for human indoor action analysis},
  author={Wang, Fei and Lv, Yizhe and Zhu, Mengdie and Ding, Han and Han, Jinsong},
  journal={Proceedings of the ACM on Interactive, Mobile, Wearable and Ubiquitous Technologies},
  volume={8},
  number={1},
  pages={1--34},
  year={2024},
  publisher={ACM New York, NY, USA}
}

@article{zhang2021comprehensive,
  title={Comprehensive mpoint: A method for 3d point cloud generation of human bodies utilizing fmcw mimo mm-wave radar},
  author={Zhang, Guangcheng and Geng, Xiaoyi and Lin, Yueh-Jaw},
  journal={Sensors},
  volume={21},
  number={19},
  pages={6455},
  year={2021},
  publisher={MDPI}
}

@inproceedings{lee2023hupr,
  title={Hupr: A benchmark for human pose estimation using millimeter wave radar},
  author={Lee, Shih-Po and Kini, Niraj Prakash and Peng, Wen-Hsiao and Ma, Ching-Wen and Hwang, Jenq-Neng},
  booktitle={Proceedings of the IEEE/CVF Winter Conference on Applications of Computer Vision},
  pages={5715--5724},
  year={2023}
}

@article{yataka2024retr,
  title={RETR: Multi-view radar detection transformer for indoor perception},
  author={Yataka, Ryoma and Cardace, Adriano and Wang, Perry and Boufounos, Petros and Takahashi, Ryuhei},
  journal={Advances in Neural Information Processing Systems},
  volume={37},
  pages={19839--19869},
  year={2024}
}

@inproceedings{fan2024diffusion,
  title={Diffusion model is a good pose estimator from 3d rf-vision},
  author={Fan, Junqiao and Yang, Jianfei and Xu, Yuecong and Xie, Lihua},
  booktitle={European Conference on Computer Vision},
  pages={1--18},
  year={2024},
  organization={Springer}
}

@article{wang2023human,
  title={Human parsing with joint learning for dynamic mmwave radar point cloud},
  author={Wang, Shuai and Cao, Dongjiang and Liu, Ruofeng and Jiang, Wenchao and Yao, Tianshun and Lu, Chris Xiaoxuan},
  journal={Proceedings of the ACM on Interactive, Mobile, Wearable and Ubiquitous Technologies},
  volume={7},
  number={1},
  pages={1--22},
  year={2023},
  publisher={ACM New York, NY, USA}
}

@article{rohling1983radar,
  title={Radar CFAR thresholding in clutter and multiple target situations},
  author={Rohling, Hermann},
  journal={IEEE transactions on aerospace and electronic systems},
  number={4},
  pages={608--621},
  year={1983},
  publisher={IEEE}
}

@inproceedings{zhao2023poseformerv2,
  title={Poseformerv2: Exploring frequency domain for efficient and robust 3d human pose estimation},
  author={Zhao, Qitao and Zheng, Ce and Liu, Mengyuan and Wang, Pichao and Chen, Chen},
  booktitle={Proceedings of the IEEE/CVF conference on computer vision and pattern recognition},
  pages={8877--8886},
  year={2023}
}

@article{zhu2024probradarm3f,
  title={ProbRadarM3F: mmWave Radar based Human Skeletal Pose Estimation with Probability Map Guided Multi-Format Feature Fusion},
  author={Zhu, Bing and He, Zixin and Xiong, Weiyi and Ding, Guanhua and Liu, Jianan and Huang, Tao and Chen, Wei and Xiang, Wei},
  journal={arXiv preprint arXiv:2405.05164},
  year={2024}
}

@inproceedings{xue2023towards,
  title={Towards generalized mmwave-based human pose estimation through signal augmentation},
  author={Xue, Hongfei and Cao, Qiming and Miao, Chenglin and Ju, Yan and Hu, Haochen and Zhang, Aidong and Su, Lu},
  booktitle={Proceedings of the 29th Annual International Conference on Mobile Computing and Networking},
  pages={1--15},
  year={2023}
}

@InProceedings{Wu_2024_CVPR,
    author    = {Wu, Guanjun and Yi, Taoran and Fang, Jiemin and Xie, Lingxi and Zhang, Xiaopeng and Wei, Wei and Liu, Wenyu and Tian, Qi and Wang, Xinggang},
    title     = {4D Gaussian Splatting for Real-Time Dynamic Scene Rendering},
    booktitle = {Proceedings of the IEEE/CVF Conference on Computer Vision and Pattern Recognition (CVPR)},
    month     = {June},
    year      = {2024},
    pages     = {20310-20320}
}

@article{landis1977measurement,
  title={The measurement of observer agreement for categorical data},
  author={Landis, J Richard and Koch, Gary G},
  journal={biometrics},
  pages={159--174},
  year={1977},
  publisher={JSTOR}
}

@inproceedings{chharia2025mv,
  title={MV-SSM: Multi-View State Space Modeling for 3D Human Pose Estimation},
  author={Chharia, Aviral and Gou, Wenbo and Dong, Haoye},
  booktitle={Proceedings of the Computer Vision and Pattern Recognition Conference},
  pages={11590--11599},
  year={2025}
}

@article{yu2023rfpose,
  title={RFPose-OT: RF-based 3D human pose estimation via optimal transport theory},
  author={Yu, Cong and Zhang, Dongheng and Wu, Zhi and Lu, Zhi and Xie, Chunyang and Hu, Yang and Chen, Yan},
  journal={Frontiers of Information Technology \& Electronic Engineering},
  volume={24},
  number={10},
  pages={1445--1457},
  year={2023},
  publisher={Springer}
}

@article{yang2023mm,
  title={Mm-fi: Multi-modal non-intrusive 4d human dataset for versatile wireless sensing},
  author={Yang, Jianfei and Huang, He and Zhou, Yunjiao and Chen, Xinyan and Xu, Yuecong and Yuan, Shenghai and Zou, Han and Lu, Chris Xiaoxuan and Xie, Lihua},
  journal={Advances in Neural Information Processing Systems},
  volume={36},
  pages={18756--18768},
  year={2023}
}

@inproceedings{ester1996density,
  title={A density-based algorithm for discovering clusters in large spatial databases with noise},
  author={Ester, Martin and Kriegel, Hans-Peter and Sander, J{\"o}rg and Xu, Xiaowei and others},
  booktitle={kdd},
  volume={96},
  number={34},
  pages={226--231},
  year={1996}
}

@misc{wu2022rfmasksimplebaselinehuman,
      title={RFMask: A Simple Baseline for Human Silhouette Segmentation with Radio Signals}, 
      author={Zhi Wu and Dongheng Zhang and Chunyang Xie and Cong Yu and Jinbo Chen and Yang Hu and Yan Chen},
      year={2022},
      eprint={2201.10175},
      archivePrefix={arXiv},
      primaryClass={cs.CV},
      url={https://arxiv.org/abs/2201.10175}, 
}

@inproceedings{xue2021mmmesh,
  title={mmMesh: Towards 3D real-time dynamic human mesh construction using millimeter-wave},
  author={Xue, Hongfei and Ju, Yan and Miao, Chenglin and Wang, Yijiang and Wang, Shiyang and Zhang, Aidong and Su, Lu},
  booktitle={Proceedings of the 19th Annual International Conference on Mobile Systems, Applications, and Services},
  pages={269--282},
  year={2021}
}

@inproceedings{kini2024transhupr,
  title={TransHuPR: Cross-View Fusion Transformer for Human Pose Estimation Using mmWave Radar},
  author={Kini, Niraj Prakash and Shiue, Ruey-Horng and Chandra, Ryan and Peng, Wen-Hsiao and Ma, Ching-Wen and Hwang, Jenq-Neng},
  booktitle={Proc. British Machine Vision Conference (BMVC 2024)},
  year={2024}
}

@article{wu2024mmhpe,
  title={mmhpe: Robust multi-scale 3d human pose estimation using a single mmwave radar},
  author={Wu, Yingxiao and Jiang, Zhongmin and Ni, Haocheng and Mao, Changlin and Zhou, Zhiyuan and Wang, Wenxiang and Han, Jianping},
  journal={IEEE Internet of Things Journal},
  year={2024},
  publisher={IEEE}
}

@article{cui2021real,
  title={Real-time short-range human posture estimation using mmWave radars and neural networks},
  author={Cui, Han and Dahnoun, Naim},
  journal={IEEE Sensors Journal},
  volume={22},
  number={1},
  pages={535--543},
  year={2021},
  publisher={IEEE}
}

@article{zhang2024single,
  title={From single-point to multi-point reflection modeling: Robust vital signs monitoring via mmwave sensing},
  author={Zhang, Duo and Zhang, Xusheng and Xie, Yaxiong and Zhang, Fusang and Yang, Hongliu and Zhang, Daqing},
  journal={IEEE Transactions on Mobile Computing},
  year={2024},
  publisher={IEEE}
}

@inproceedings{rahman2024mmvr,
  title={MMVR: Millimeter-Wave Multi-view Radar Dataset and Benchmark for Indoor Perception},
  author={Rahman, M Mahbubur and Yataka, Ryoma and Kato, Sorachi and Wang, Pu and Li, Peizhao and Cardace, Adriano and Boufounos, Petros},
  booktitle={European Conference on Computer Vision},
  pages={306--322},
  year={2024},
  organization={Springer}
}

\appendix
\section{Appendix}

\subsection{Velocity Initialization for Doppler-Deficient Datasets}
\label{appendix:velocity_approx}

For datasets where explicit Doppler dimensions are unavailable due to limited chirp counts or different FFT processing algorithms \cite{rahman2024mmvr}, we approximate velocity from spatial intensity gradients inspired by optical flow principles \cite{alfarano2024estimating}. The core insight is that temporal intensity changes in the Heatmap correlate with target motion: regions with high gradient magnitude typically correspond to moving objects.

Given consecutive Heatmaps $H_t$ and $H_{t+1}$ at times $t$ and $t+1$, we compute the spatial gradient magnitude for each candidate peak location $(n_r, n_a)$:
\begin{equation}
\small
g_{t}(n_r, n_a) = \sqrt{\left(\frac{\partial H_t}{\partial n_r}\right)^2 + \left(\frac{\partial H_t}{\partial n_a}\right)^2},
\end{equation}
where partial derivatives are approximated via finite differences. The radial velocity is then estimated as:
\begin{equation}
\small
\hat{v}_r = \gamma \cdot g_t(n_r, n_a),
\end{equation}
where $\gamma$ is a scaling factor calibrated to typical human motion speeds (empirically set to $\gamma=0.5$ m/s based on walking velocity statistics). This gradient-based solution provides coarse velocity priors that are subsequently refined under electromagnetic reconstruction constraints during optimization (Sec. 4.3).

We validate this approximation on the MMVR dataset's single-person subset by comparing gradient-based estimates against ground-truth Doppler measurements. On 1000 randomly sampled frames, the correlation coefficient between $\hat{v}_r$ and true radial velocities is 0.67, indicating moderate agreement sufficient for initialization purposes.

\subsection{Person Counting Module Architecture}
\label{appendix:person_counting}

The person counting module consists of three stages: CFAR-based peak detection, DBSCAN clustering, and MLP classification.

\textbf{CFAR Peak Detection.} We adopt ETCM-CFAR (Enhanced Temporal Correlation Matrix CFAR) \cite{yang2025spatial}, which improves upon standard CFAR by incorporating temporal consistency across consecutive frames. For each cell $(n_r, n_a, n_d)$ (range-angle-Doppler)in the Heatmap, we compute ETCM-CFAR:
\begin{equation}
\small
\text{threshold}(n_r, n_a, n_d) = \mu_{\text{ref}} + \beta \sigma_{\text{ref}},
\end{equation}
where $\mu_{\text{ref}}$ and $\sigma_{\text{ref}}$ are the mean and standard deviation of reference cells (excluding guard cells), and $\beta=3.0$ is the detection sensitivity parameter. Peaks exceeding the threshold are retained as candidate detections.

\textbf{DBSCAN Clustering.} Detected peaks are projected to Cartesian coordinates $(x, y, z)$ and clustered using DBSCAN \cite{ester1996density} with parameters: $\epsilon=0.3$m (neighborhood radius) and $\text{minPts}=3$ (minimum cluster size). Each cluster corresponds to a hypothesized person.

\textbf{MLP Classifier.} Cluster statistics are fed into a 3-layer MLP to refine person counts:
\begin{itemize} 
\item \textbf{Input features}: Output of DBSCAN, including cluster centroid $(x, y, z)$, spatial extent $(\Delta x, \Delta y, \Delta z)$, mean intensity $\mu_I$, intensity variance $\sigma_I^2$, point count $N_{\text{pts}}$, and temporal consistency score $C_t$.
\item \textbf{Architecture}: FC(12→64)→ReLU→Dropout(0.3)→FC(64→32)→ReLU→FC(32→5), where the output layer predicts person count $\in \{0,1,2,3,4+\}$.
\item \textbf{Training}: Cross-entropy loss on 5000 labeled Heatmap samples from MMVR dataset, Adam optimizer ($\text{lr}=10^{-3}$), 50 epochs.
\end{itemize}

\textbf{Performance.} On MMVR, the person counting module achieves 94.2\% accuracy for exact count prediction and 98.7\% accuracy within ±1 person. Failure cases primarily occur during rapid person entry/exit transitions.

\subsection{Count-Matched Evaluation Protocol: Detailed Analysis}
\label{appendix:count_matched}

Multi-person HPE evaluation in mmWave datasets faces systematic annotation challenges due to vision-based labeling limitations in experimental datasets: MMVR \cite{rahman2024mmvr}, HuPR \cite{lee2023hupr}, and XRF55 \cite{wang2024xrf55}. This appendix provides detailed justification for the count-matched evaluation protocol adopted in Sec. 5.3.

\subsubsection{Annotation Error Mechanisms}

\begin{figure}[h]
  \centering
  \includegraphics[width=0.8\linewidth]{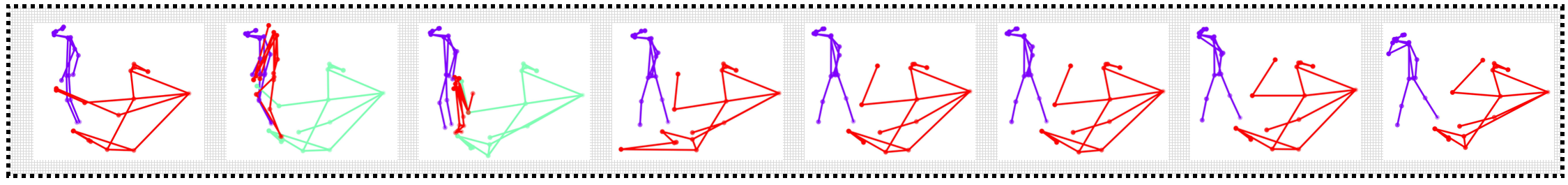}
  \caption{Representative annotation error in MMVR dataset.}
  \label{fig:vision_error}
\end{figure}

Vision-based SOTA estimation systems \cite{chharia2025mv} achieve <95\% average precision in multi-person scenarios. Errors manifest in three primary modes:

\begin{enumerate} 
\item \textbf{Skeletal identity misassignment}: When two individuals are spatially proximate, pose estimators occasionally assign joints from one person to another's skeleton (Fig.~\ref{fig:vision_error}). This creates anatomically implausible skeletons (e.g., limbs spanning <2cm) that violate biomechanical constraints.

\item \textbf{Tracking failures during occlusions}: Person re-identification across frames fails when individuals temporarily occlude each other, causing identity swaps in annotation tracks.

\item \textbf{Detection failures}: Vision systems fail to detect all persons when lighting conditions are poor or individuals are partially occluded, resulting in count mismatches between true occupancy and annotated persons.
\end{enumerate}

\subsubsection{Correlation Between Count Mismatches and Annotation Quality}

We hypothesize that count-mismatch events (where predicted count $N_{\text{pred}} \neq$ annotated count $N_{\text{GT}}$) correlate with elevated annotation error rates. To validate this, two domain experts with 5+ years of HPE research experience independently reviewed 200 randomly sampled count-mismatch frames from MMVR \cite{rahman2024mmvr} (100 frames) and XRF55 \cite{wang2024xrf55} (100 frames).

\textbf{Review protocol}: For each frame, experts examined synchronized RGB video and Heatmap visualizations, categorizing the mismatch cause into three classes:
\begin{itemize} 
\item \textbf{Annotation Error}: Vision labels are incorrect (wrong count or mis-attributed joints).
\item \textbf{Detection Error}: Our person counting module prediction is incorrect.
\item \textbf{Both Incorrect}: Both systems produce errors.
\end{itemize}

\textbf{Inter-rater reliability}: Cohen's kappa coefficient $\kappa=0.91$ indicates high agreement between experts (substantial agreement: $\kappa>0.80$ \cite{landis1977measurement}). Disagreements were resolved through consensus discussion.

\textbf{Results}: Table~\ref{tab:verification_detailed} summarizes findings. Annotation errors account for 9\% of count-mismatch frames, exceeding detection errors (2\%) by 4.5×. The remaining 87\% are true negatives where both systems are correct and mismatches arise from transient tracking variations.

\begin{table}[h]
\centering
\begin{adjustbox}{width=0.5\textwidth,center}
\setlength{\tabcolsep}{1pt}
\begin{tabular}{lrrr}
\hline
\textbf{Error Source} & \textbf{MMVR} & \textbf{XRF55} & \textbf{Combined} \\ \hline
\rowcolor[HTML]{FFE6E6}
Annotation Error & 11/100 (11\%) & 7/100 (7\%) & \textbf{18/200 (9\%)} \\ 
Detection Error & 3/100 (3\%) & 1/100 (1\%) & 4/200 (2\%) \\ 
Both Incorrect & 3/100 (3\%) & 1/100 (1\%) & 4/200 (2\%) \\ 
True Negative & 83/100 (83\%) & 91/100 (91\%) & 174/200 (87\%) \\ \hline
\end{tabular}
\end{adjustbox}
\caption{Manual verification of count-mismatch frames. Annotation errors (9\%) dominate over detection errors (2\%), validating count-matched filtering as an effective annotation quality control mechanism.}
\label{tab:verification_detailed}
\end{table}

\subsubsection{Count-Matched Protocol Definition}

Based on these findings, we adopt the following protocol:

\textbf{Filtering criterion}: Evaluate only frames where:
\begin{equation}
\small
N_{\text{pred}} = N_{\text{GT}} = N_{\text{verified}},
\end{equation}
where $N_{\text{pred}}$ is the person counting module output, $N_{\text{GT}}$ is the dataset annotation, and $N_{\text{verified}}$ is manual verification (for validation subset).

\textbf{Coverage}: Count-matched filtering retains 91-95\% of original multi-person frames across datasets:
\begin{itemize} 
\item MMVR: 93.2\% retention
\item XRF55: 94.8\% retention
\item HuPR: Limited multi-person data (excluded from multi-person evaluation)
\end{itemize}

\textbf{Impact on HPE models}: This filtering protocol affects all supervised HPE models equally, ensuring fair comparison. PPPR benefits from skeletal topology constraints that inherently filter anatomically implausible annotations during optimization, providing additional robustness.

\subsection{Additional Implementation Details}
\label{appendix:implementation}

\textbf{PPPR optimization hyperparameters}:
\begin{itemize}
\item Optimizer: Adam with learning rate $\alpha=10^{-3}$, $\beta_1=0.9$, $\beta_2=0.999$
\item Iterations: 100 per frame (converges in 50-80 iterations typically)
\item Loss weights: $w_{\mathrm{EM}}=0.5$, $w_{\mathrm{kine}}=0.5$
\item Bone length tolerance: $\pm$5cm from reference lengths
\item Joint angle limits: $\theta_{\max}=170°$ (allows near-full extension)
\end{itemize}

\textbf{Training configurations}:
\begin{itemize} 
\item Batch size: 32 (RETR, HuprModel), 16 (mmDiff, PoseformerV2)
\item Learning rate: $10^{-4}$ with cosine annealing
\item Training epochs: 200 (early stopping with patience=20)
\item Data augmentation: Random rotation (±15°), translation (±0.1m), scaling (0.9-1.1×)
\end{itemize}

\textbf{Computational requirements}:
\begin{itemize} 
\item PPPR preparation: 3.2ms/frame (single-person), 4.5-5.9ms/frame (multi-person)
\item HPE model inference: 8-15ms/frame (architecture-dependent)
\item Total latency: 11-21ms/frame (48-91 FPS)
\item Memory footprint: PPPR parameters 2.8KB/person, Heatmap 256KB
\end{itemize}

\subsection{\rev{End-to-end runtime and online feasibility}}
\label{appendix:end_to_end_runtime}

\rev{This appendix reports cumulative computational costs beyond backbone inference, including initialization, differentiable radar simulation, and iterative optimization. We provide a latency breakdown and an online warm-start setting under fixed iteration budgets.}

\begin{table}[h]
\centering
\setlength{\tabcolsep}{2pt}
\begin{adjustbox}{width=0.8\linewidth,center}
\begin{tabular}{lcccccc}
\hline
\textbf{Setting} & \textbf{Init (ms)} & \textbf{Sim (ms)} & \textbf{Opt/iter (ms)} & \textbf{\#iters} & \textbf{Backbone (ms)} & \textbf{Total / FPS} \\
\hline
Offline (per-frame init) & \textbf{1.20} & \textbf{0.25} & \textbf{0.03} & \textbf{80} & \textbf{11.00} & \textbf{14.85 / 67.3} \\
Online (warm-start)      & \textbf{0.25} & \textbf{0.25} & \textbf{0.03} & \textbf{25} & \textbf{11.00} & \textbf{12.25 / 81.6} \\
\hline
\end{tabular}
\end{adjustbox}
\caption{\rev{End-to-end latency breakdown of the full PPPR pipeline. ``Init'' includes peak extraction/clustering and parameter initialization; ``Sim'' includes differentiable radar forward simulation; ``Opt/iter'' is the per-iteration cost for optimizing Eq.~\ref{eq:total_loss} under a fixed iteration budget.}}
\label{tab:end_to_end_latency}
\end{table}

\rev{Table~\ref{tab:end_to_end_latency} decomposes end-to-end latency into initialization, radar simulation, iterative optimization, and backbone inference. In the offline setting, PPPR preparation (Init + Sim + Opt) totals $1.20 + 0.25 + 80 \times 0.03 = 3.85$ ms per frame, while backbone inference accounts for 11.00 ms. The per-iteration optimization cost is 0.03 ms/iter, so a budget of 80 iterations adds 2.40 ms.}

\rev{In the online warm-start setting, the iteration budget can be reduced (e.g., to 25 iterations) because consecutive frames provide a strong initial estimate. The optimization term therefore decreases to $25 \times 0.03 = 0.75$ ms, yielding 12.25 ms per frame (81.6 FPS). In these settings, the iteration budget provides a direct accuracy--latency control knob, while backbone inference remains the dominant term in end-to-end latency.}

\begin{figure}[h]
\centering
\includegraphics[width=0.5\linewidth]{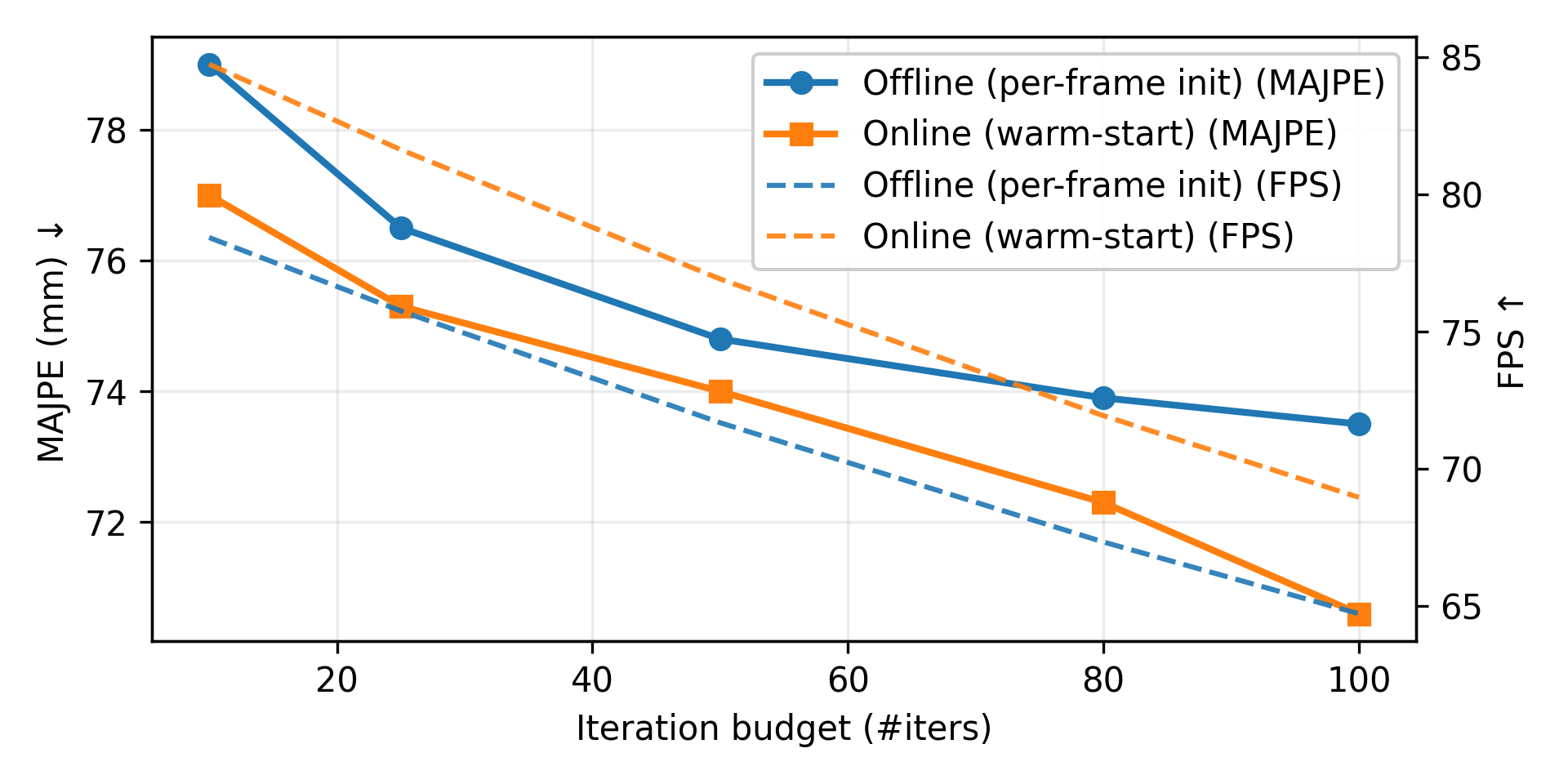}
\caption{\rev{Speed--accuracy trade-off under fixed iteration budgets. The plot reports MAJPE and FPS versus iteration budget for per-frame initialization and online warm-start settings.}}
\label{fig:speed_accuracy_tradeoff}
\end{figure}

\rev{Figure~\ref{fig:speed_accuracy_tradeoff} summarizes the speed--accuracy trade-off of PPPR under fixed iteration budgets. As the iteration budget increases, MAJPE improves rapidly at small budgets and then exhibits diminishing returns, indicating that most improvements occur within a moderate number of iterations. Importantly, the online warm-start setting achieves comparable accuracy with fewer iterations than per-frame initialization, suggesting that temporal continuity provides a favorable initialization and reduces the per-frame optimization workload.}

\rev{From the runtime perspective, the FPS curve is largely governed by the backbone inference time, while the PPPR optimization cost scales approximately linearly with the iteration budget due to the small per-iteration cost. Consequently, reducing the iteration budget (especially under warm-start) yields a near-linear runtime gain, with a modest change in accuracy beyond the knee region. Overall, these results suggest that the iteration budget provides an explicit accuracy--latency trade-off for deployment.}

\subsection{\rev{MLP architecture for the MLP+PPPR baseline}}
\label{appendix:mlp_pppr_arch}

\rev{For the lightweight ``MLP+PPPR'' configuration in Table~\ref{mlp}, we use a small feed-forward network that regresses 3D joint coordinates from PPPR parameters. The input is constructed by concatenating per-joint PPPR parameters and applying mean pooling over joints, yielding a fixed-length feature vector of dimension \textbf{48}. We use a 3-layer MLP with dimensions \textbf{48} $\rightarrow$ \textbf{2048} $\rightarrow$ \textbf{1792} $ \rightarrow$ $3N_j$, with ReLU activations and dropout rate \textbf{0.30}. This design keeps the baseline compact while isolating the effect of the PPPR representation.}

\subsection{\rev{Acronym List}}
\label{appendix:acronyms}

\rev{To improve readability, Table~\ref{tab:acronyms} summarizes the key acronyms used throughout the paper.}
\begin{table}[h]
\centering
\footnotesize
\renewcommand{\arraystretch}{0.8}
\begin{adjustbox}{width=\textwidth,center}
\setlength{\tabcolsep}{4pt}
\begin{tabular}{c p{0.27\textwidth} c | c p{0.27\textwidth} c}
\hline
\textbf{Acronym} & \textbf{Full name} & \textbf{First} & \textbf{Acronym} & \textbf{Full name} & \textbf{First} \\ \hline
HPE & Human Pose Estimation & Sec. 1 & mmWave & millimeter-wave & Sec. 1 \\
PPPR & Person Parametric Physics-informed Representation & Sec. 1 & MHP & MmWave Human Parameterization & Sec. 1 \\
FFT & Fast Fourier Transform & Sec. 2.1 & FMCW & Frequency-Modulated Continuous Wave & Sec. 2.1 \\
CFAR & Constant False Alarm Rate & Sec. 1 & PC & Point Cloud & Sec. 1 \\
Tx--Rx & transmit--receive (antenna array) & Sec. 2.1 & RCS & Radar Cross Section & Sec. 4.1 \\
IoU & Intersection-over-Union & Sec. 4.3 & GS & Gaussian Splatting & Sec. 2.2 \\
MAJPE & Mean Absolute Joint Position Error & Sec. 5.1 & PA-MAJPE & Procrustes-Aligned MAJPE & Sec. 5.1 \\
DBSCAN & Density-Based Spatial Clustering of Applications with Noise & Sec. 4.4 & MLP & Multi-Layer Perceptron & Sec. 4.4 \\
PRF & Pulse Repetition Frequency & Sec. 4.1 & SNR & Signal-to-Noise Ratio & Sec. 5.4 \\
\hline
\end{tabular}
\end{adjustbox}
\caption{\rev{Key acronyms used throughout the paper.}}
\label{tab:acronyms}
\end{table}

\subsection{Symbol Table}
\label{appendix:symbols}

Table~\ref{tab:symbols} summarizes major mathematical symbols used throughout this paper.

\begin{table}[h]
\centering
\footnotesize
\renewcommand{\arraystretch}{0.8}
\begin{adjustbox}{width=\textwidth,center}
\setlength{\tabcolsep}{1pt}
\begin{tabular}{c p{0.26\textwidth} c | c p{0.26\textwidth} c}
\hline
\textbf{Symbol} & \textbf{Description} & \textbf{First} & \textbf{Symbol} & \textbf{Description} & \textbf{First} \\ \hline
\multicolumn{6}{c}{\textit{Radar Signal Processing}} \\ \hline
$\lambda$ & Electromagnetic wavelength & Sec. 2.1 & $c$ & Speed of light & Sec. 2.1 \\
$r$ & Range (distance to target) & Sec. 2.1 & $\theta_{\mathrm{az}}, \theta_{\mathrm{el}}$ & Azimuth and elevation angles & Sec. 2.1 \\
$l_{\mathrm{az}}, l_{\mathrm{el}}$ & Antenna element spacing (azimuth, elevation) & Sec. 2.1 & $\Delta\Phi_{\mathrm{az}}, \Delta\Phi_{\mathrm{el}}$ & Phase differences (azimuth, elevation) & Sec. 2.1 \\
$v_r$ & Radial velocity & Sec. 2.1 & $\Delta f_d$ & Doppler frequency shift & Sec. 2.1 \\
$B, T_c, S$ & Chirp bandwidth, duration, slope & Sec. 2.1 & $f_{\mathrm{beat}}$ & Beat frequency & Sec. 2.1 \\
\hline
\multicolumn{6}{c}{\textit{Data Representations}} \\ \hline
$H \in \mathbb{R}^{N_r \times N_a}$ & Heatmap (range $\times$ azimuth) & Sec. 2.1 & $N_r, N_a, N_e$ & Number of range, azimuth, elevation bins & Sec. 2.1 \\
$H_{\mathrm{ori}}$ & Observed Heatmap (input) & Sec. 4.1 & $H_{\mathrm{sim}}$ & Simulated Heatmap & Sec. 4.2 \\
\hline
\multicolumn{6}{c}{\textit{PPPR Parameters}} \\ \hline
$N_j$ & Number of joints & Sec. 4.1 & $\Theta_j$ & Parameter set for joint $j$ & Sec. 4.1 \\
$\mathbf{p}_j \in \mathbb{R}^3$ & Position (3D centroid) & Sec. 4.1 & $\mathbf{s}_j \in \mathbb{R}^3$ & Scale (anisotropic extent) & Sec. 4.1 \\
$\mathbf{q}_j \in \mathbb{R}^4$ & Rotation (quaternion) & Sec. 4.1 & $\mathbf{v}_j \in \mathbb{R}^3$ & Velocity (Cartesian) & Sec. 4.1 \\
$\beta_j$ & Opacity (radar cross-section) & Sec. 4.1 & $\boldsymbol{\omega}_j \in \mathbb{R}^{N_d}$ & Doppler features & Sec. 4.1 \\
$\mathbf{R}_j \in SO(3)$ & Rotation matrix & Sec. 4.1 & $\boldsymbol{\Sigma}_j$ & Covariance matrix & Sec. 4.1 \\
$\mathcal{R}_j(\mathbf{x})$ & Radar return contribution & Sec. 4.1 &  &  &  \\
\hline
\multicolumn{6}{c}{\textit{Radar Simulation}} \\ \hline
$\mathbf{i}$ & Imaginary unit $\sqrt{-1}$ & Sec. 4.2 & $d_j$ & Range distance $\|\mathbf{p}_j\|$ & Sec. 4.2 \\
$M_{\mathrm{atten}}, M_{\mathrm{range}}, M_{\mathrm{Dopp}}, M_{\mathrm{angle}}$ & Signal modulation operators & Sec. 4.2 & $T_{\mathrm{frame}}$ & Frame duration (coherent processing interval) & Sec. 4.2 \\
\hline
\multicolumn{6}{c}{\textit{Optimization Constraints}} \\ \hline
$\mathcal{E}$ & Set of skeletal edges (bones) & Sec. 4.3 & $\mathcal{A}$ & Set of joint angle triplets & Sec. 4.3 \\
$\ell_{mn}$ & Bone length between joints $m$ and $n$ & Sec. 4.3 & $\hat{\mathbf{b}}_{mn}$ & Unit bone direction vector & Sec. 4.3 \\
$\theta_{mno}$ & Joint angle at vertex $n$ & Sec. 4.3 & $\theta_{\max}$ & Maximum allowed joint angle & Sec. 4.3 \\
$\mathcal{B}_{\mathrm{sim}}, \mathcal{B}_{\mathrm{obs}}$ & Binary masks for IoU computation & Sec. 4.3 & $\tau_{\mathrm{pct}}$ & Percentile threshold for masking & Sec. 4.3 \\
$w_{\mathrm{EM}}, w_{\mathrm{kine}}$ & Loss function weights & Sec. 4.3 &  &  &  \\
\hline
\multicolumn{6}{c}{\textit{Multi-Person Extension}} \\ \hline
$N_{\mathrm{person}}$ & Number of persons & Sec. 4.4 & $s, t$ & Person indices & Sec. 4.4 \\
$\mathbf{c}_s$ & Skeleton centroid for person $s$ & Sec. 4.4 & $d_{\mathrm{sep}}$ & Minimum centroid separation distance & Sec. 4.4 \\
$d_{\mathrm{joint}}$ & Minimum joint-level separation distance & Sec. 4.4 &  &  &  \\
\hline
\end{tabular}
\end{adjustbox}
\caption{Major symbols and notation used throughout the paper.}
\label{tab:symbols}
\end{table}

\end{document}